\documentclass[sigconf,screen]{acmart}

\AtBeginDocument{%
  \providecommand\BibTeX{{%
    \normalfont B\kern-0.5em{\scshape i\kern-0.25em b}\kern-0.8em\TeX}}}

\newcommand{\ignore}[1]{}

\usepackage{algorithm}
\usepackage{algpseudocode}
\usepackage{graphicx}
\usepackage{subcaption}

\usepackage{ulem}
\usepackage[switch]{lineno}
\usepackage{hyperref}
            
\usepackage{listings, xcolor}

\definecolor{verylightgray}{rgb}{.97,.97,.97}
\lstdefinelanguage{Solidity}{
  keywords=[1]{anonymous, assembly, assert, balance, break, call, callcode, case, catch, class, constant, continue, constructor, contract, debugger, default, delegatecall, delete, do, else, emit, event, experimental, export, external, false, finally, for, function, gas, if, implements, import, in, indexed, instanceof, interface, internal, is, length, library, log0, log1, log2, log3, log4, memory, modifier, new, payable, pragma, private, protected, public, pure, push, require, return, returns, revert, selfdestruct, send, solidity, storage, struct, suicide, super, switch, then, this, throw, transfer, true, try, typeof, using, value, view, while, with, addmod, ecrecover, keccak256, mulmod, ripemd160, sha256, sha3}, 
  keywordstyle=[1]\color{blue}\bfseries,
  keywords=[2]{address, bool, byte, bytes, bytes1, bytes2, bytes3, bytes4, bytes5, bytes6, bytes7, bytes8, bytes9, bytes10, bytes11, bytes12, bytes13, bytes14, bytes15, bytes16, bytes17, bytes18, bytes19, bytes20, bytes21, bytes22, bytes23, bytes24, bytes25, bytes26, bytes27, bytes28, bytes29, bytes30, bytes31, bytes32, enum, int, int8, int16, int24, int32, int40, int48, int56, int64, int72, int80, int88, int96, int104, int112, int120, int128, int136, int144, int152, int160, int168, int176, int184, int192, int200, int208, int216, int224, int232, int240, int248, int256, mapping, string, uint, uint8, uint16, uint24, uint32, uint40, uint48, uint56, uint64, uint72, uint80, uint88, uint96, uint104, uint112, uint120, uint128, uint136, uint144, uint152, uint160, uint168, uint176, uint184, uint192, uint200, uint208, uint216, uint224, uint232, uint240, uint248, uint256, var, void, ether, finney, szabo, wei, days, hours, minutes, seconds, weeks, years},  
  keywordstyle=[2]\color{teal}\bfseries,
  keywords=[3]{block, blockhash, coinbase, difficulty, gaslimit, number, timestamp, msg, data, gas, sender, sig, value, now, tx, gasprice, origin},  
  keywordstyle=[3]\color{violet}\bfseries,
  identifierstyle=\color{black},
  sensitive=false,
  comment=[l]{//},
  morecomment=[s]{/*}{*/},
  commentstyle=\color{gray}\ttfamily,
  stringstyle=\color{red}\ttfamily,
  morestring=[b]',
  morestring=[b]"
}
\usepackage{colortbl}

\usepackage{pifont}
\usepackage{diagbox}
\usepackage{tcolorbox}
\usepackage{color}
\usepackage{microtype}
\usepackage{balance}
\usepackage{oplotsymbl}
\usepackage{siunitx}
\usepackage{array,framed}
 \usepackage{
   color,
   float,
   epsfig,
   wrapfig,
   graphics,
   graphicx,
 }
\usepackage{setspace}
\usepackage{amsfonts}
\usepackage{latexsym,fancyhdr,url}
\usepackage{enumerate}
\usepackage{algpseudocode}
\usepackage{graphics}
\usepackage{xparse} 
\usepackage{xspace}
\usepackage{multirow}
\usepackage{appendix}
\usepackage{microtype}
\usepackage{enumitem}

\usepackage{soul}
\soulregister\cite7 
\soulregister\ref7 

\usepackage{color}

\usepackage{fancyhdr}

\newcommand{\revise}[1]{{#1}}

\newcommand{\later}[1]{}
\newcommand{\Our}{\textsc{CodeFast}\xspace}

\newcommand{\GenGuard}{GenGuard\xspace} 
\newcommand{\CLSeven}{CodeLlama-7B\xspace}

\newcommand{\CLSevenOurs}{CodeLlama-7B\textsubscript{\Our}\xspace}
\newcommand{\CLThirteen}{CodeLlama-13B\xspace}
\newcommand{\CLThirteenOurs}{CodeLlama-13B\textsubscript{\Our}\xspace}

\newcommand{\CLThirtyFour}{CodeLlama-34B\xspace}
\newcommand{\CLThirtyFourOurs}{CodeLlama-34B\textsubscript{\Our}\xspace}

\newcommand{\StarCoder}{StarCoder-15.5B\xspace}
\newcommand{\StarCoderOurs}{StarCoder-15.5B\textsubscript{\Our}\xspace}

\newcommand{\PhindCL}{Phind-CodeLlama\xspace}
\newcommand{\PhindCLOurs}{Phind-CodeLlama\textsubscript{\Our}\xspace}

\newcommand{\Fig}{Fig.\xspace}

\newcommand{\Sec}{\S\xspace} 

\newcommand{\secmargin}{\vspace{-2mm}} 
\newcommand{\figmargin}{\vspace{-4mm}} 
\newcommand{\tabmargin}{\vspace{-3mm}} 
\newcommand{\eqmargin}{\vspace{-1mm}} 

\newcommand{\boxmargin}{1mm}
\tcbuselibrary{most, skins, breakable, theorems}
\newtcolorbox{myboxa}[2][]{
    colback=gray!10!white,
    colframe=black, enhanced,
    attach boxed title to top left={yshift=-2mm,xshift=5mm},
    title=#2,#1
}
\newtcolorbox{myboxb}[2][]{
    boxsep=1pt,
    left = \boxmargin, right = \boxmargin, top = \boxmargin, bottom = \boxmargin,
    title={#2},#1
}
\newtcolorbox{myboxc}{
    colback=gray!15!white,
    arc = 0pt, outer arc = 0pt,
    boxsep=0pt, left = 3pt, right = 0pt, top = 0pt, bottom = 0pt, 
    leftrule=3pt, bottomrule=0pt,toprule=0pt, rightrule=0pt,
    left = \boxmargin, right = \boxmargin, top = \boxmargin, bottom = \boxmargin
}

\setcopyright{none}
\renewcommand\footnotetextcopyrightpermission[1]{}
\begin{document}

\title{When to Stop? Towards Efficient Code Generation in LLMs with Excess Token Prevention}

\author[L. Guo]{Lianghong Guo}
\orcid{0009-0001-0943-5049}
\affiliation{%
  \institution{Sun Yat-sen University}
  \city{Zhuhai}
  \country{China}
}
\email{guolh8@mail2.sysu.edu.cn}

\author[Y. Wang]{Yanlin Wang}
\orcid{0000-0001-7761-7269}
\authornote{Yanlin Wang is the corresponding author.}

\affiliation{%
  \institution{Sun Yat-sen University}
  \city{Zhuhai}
  \country{China}
}
\email{wangylin36@mail.sysu.edu.cn}

\author[E. Shi]{Ensheng Shi }
\orcid{0000-0002-5543-2025}
\affiliation{%
  \institution{Xi'an Jiaotong University}
  \city{Xi'an}
  \country{China}
}
\email{s1530129650@stu.xjtu.edu.cn}

\author[W. Zhong]{Wanjun Zhong}
\orcid{0009-0007-2236-228X}
\affiliation{%
   \institution{Sun Yat-sen University}
  \city{Guangzhou}
  \country{China}
}
\email{zhongwj25@mail2.sysu.edu.com}

\author[H. Zhang]{Hongyu Zhang}
\orcid{0000-0002-3063-9425}
\affiliation{%
  \institution{Chongqing University}
  \city{Chongqing}
  \country{China}
}
\email{hyzhang@cqu.edu.cn}

\author[J. Chen]{Jiachi Chen }
\orcid{0000-0002-0192-9992}
\affiliation{%
  \institution{Sun Yat-sen University}
  \city{Zhuhai}
  \country{China}
}
\email{chenjch86@mail.sysu.edu.cn}

\author[R. Zhang]{Ruikai Zhang}
\orcid{0000-0001-8929-628X}
\affiliation{%
  \institution{Huawei Cloud Computing Technologies Co., Ltd.}
  \city{Shenzhen}
  \country{China}
}
\email{zhangruikai1@huawei.com}

\author[Y. Ma]{Yuchi Ma}
\orcid{0009-0002-3304-1389}
\affiliation{%
  \institution{Huawei Cloud Computing Technologies Co., Ltd.}
  \city{Shenzhen}
  \country{China}
}
\email{mayuchi1@huawei.com}

\author[Z. Zheng]{Zibin Zheng}
\orcid{0000-0002-7878-4330}
\affiliation{%
  \institution{Sun Yat-sen University}
  \city{Zhuhai}
  \country{China}
}
\email{zhzibin@mail.sysu.edu.cn}

\begin{abstract}

Code generation aims to automatically generate code snippets that meet given natural language requirements and plays an important role in software development. Although Code LLMs have shown excellent performance in this domain, their long generation time poses a signification limitation in practice use. In this paper, we first conduct an in-depth preliminary study with different Code LLMs on code generation tasks and identify a significant efficiency issue, i.e., \textit{continual generation of excess tokens}. It harms the developer productivity and leads to huge computational wastes. To address it, we introduce \textbf{\Our}, an inference acceleration approach for Code LLMs on code generation. The key idea of \Our is to terminate the inference process in time when unnecessary excess tokens are detected. First, we propose an automatic data construction framework to obtain training data. Then, we train a unified lightweight model \GenGuard applicable to multiple programming languages to predict whether to terminate inference at the current step. Finally, we enhance Code LLM with \GenGuard to accelerate its inference in code generation tasks. We conduct extensive experiments with \Our on five representative Code LLMs across four widely used code generation datasets. Experimental results show that (1) \Our can significantly improve the inference speed of various Code LLMs in code generation, ranging form 34\% to 452\%, without compromising the quality of generated code. (2) \Our is stable across different parameter settings and can generalize to untrained datasets. Our code and data are available at \color{blue}{\url{https://github.com/DeepSoftwareAnalytics/CodeFast}}.

\end{abstract}

\maketitle

\secmargin
\section{Introduction}
\label{sec:intro}
Code generation aims to automatically generate code snippets that meet given natural language requirements and plays an important role in software development~\cite{openai2021codex,zhang2023draft,radford2018improving,zan2023large,olausson2023demystifying,le2022coderl,muennighoff2023octopack,zan2023private,huang2023enhancing,jiang2021exploring,agrawal2023guiding,yu2023codereval,roziere2023code,luo2023wizardcoder,phind2023codellama,li2023starcoder,li2023think,liu2023refining,zhang2023repocoder,sun2023don,zheng2023survey,zhang2023survey,jain2023llm,bairi2023codeplan,zhang2023planning,ni2023lever,chen2023improving,li2023structured,yadav2023exploring,li2023large,li2023towards,zhang2023self,zhu2023improving,wang2021code,shi2023towards,shi2023sotana,nie2023unveiling}. 
Recently, an increasing number of 
large language model for code (Code LLMs), such as Codex~\cite{openai2021codex},
StarCoder~\cite{li2023starcoder}, and Code Llama~\cite{roziere2023code}, have achieved remarkable performance in code generation~\cite{zhang2023survey,zheng2023survey}. Among them, Code Llama~\cite{roziere2023code} stands out as a powerful open-source model in code generation tasks. In particular, models fine-tuned based on Code Llama, such as WizardCoder~\cite{luo2023wizardcoder} and Phind CodeLlama~\cite{phind2023codellama}, have demonstrated performance surpassing GPT-4~\cite{openai2023gpt4} in code generation.

\begin{figure}[t]
    \centering
    \includegraphics[width=0.9999\linewidth]{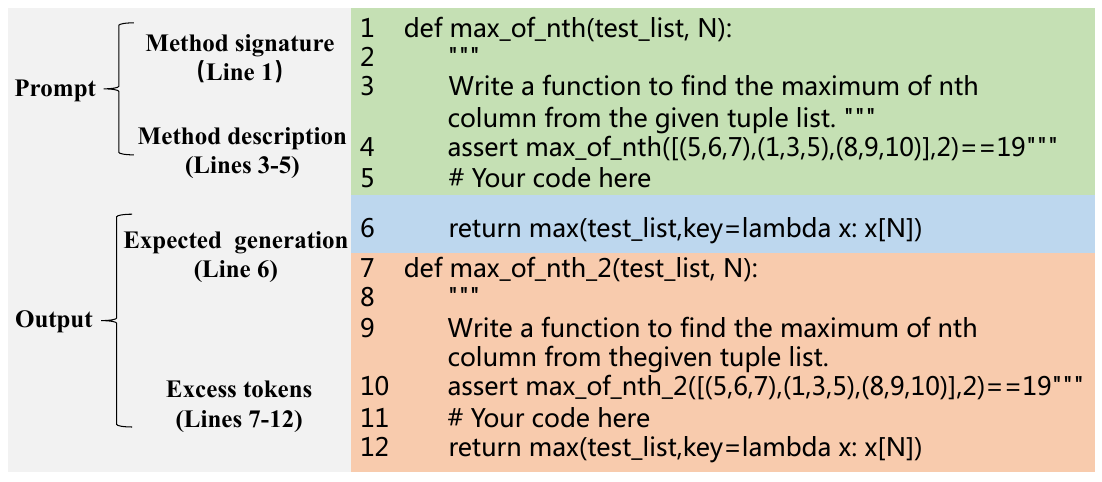}
    \vspace{-1mm}
    \caption{A motivating example of code generated from CodeLlama-7B.}
    \label{fig:motivating_example}
\end{figure}

While Code LLMs~\cite{phind2023codellama,roziere2023code,li2023starcoder,luo2023wizardcoder} have shown impressive code generation capabilities, they exhibit inefficiencies, such as continuing to generate unnecessary excess code snippets even after the expected code is generated. For example, \Fig~\ref{fig:motivating_example} presents a code snippet generated by \CLSeven~\cite{roziere2023code}, including two parts: prompt (lines 1-5) and output (lines 6-12). Generally, we provide the prompt, including method signature (lines 1) and method description (lines 3-5), as the input to \CLSeven. The model generates the expected code fragment in line 6, which already meets the specified requirement. However, the model fails to terminate its inference here and continues to generate excess code (lines 7-12) until reaching the predefined maximum token generation length. The unnecessary generation of excess code not only wastes computational resources but also leads to significant energy consumption. Moreover, developers have to check the generated results and remove this excess code, harming the development productivity. Therefore, \textit{there is a clear need to understand to what extent Code LLMs are affected by this excess token generation issue and how to effectively handle it.}

In this paper, we conduct extensive preliminary studies on five Code LLMs across four popular code generation benchmarks to study to what extent Code LLMs are affected by the \textit{excess token generation} 
issue and then introduce \textbf{\Our}, a straightforward and effective inference acceleration approach for Code LLMs to address it. 

\begin{figure}[t]
    \centering
 
    \includegraphics[width=\linewidth]{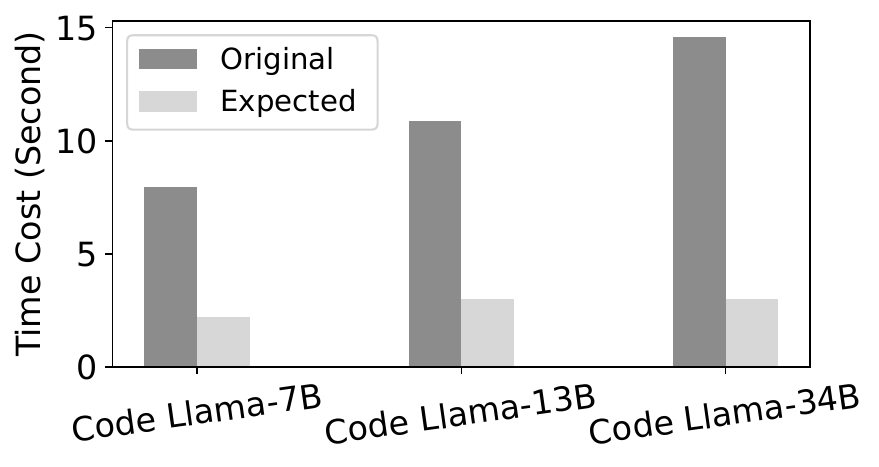}
    \vspace{-5mm}
    \caption{Experimental results of Code LLMs on the MBGP dataset: a comparison between original and expected scenarios.}
    \label{fig:comparison_of_different_scenarios}
    \figmargin
    \vspace{7mm}
\end{figure}

In preliminary study, we first randomly select 100 problems from the training sets of the studied datasets (MBPP~\cite{austin2021program}, MBJSP~\cite{athiwaratkun2022multi}, MBGP~\cite{athiwaratkun2022multi}, and MBCPP~\cite{athiwaratkun2022multi}). Next, we build prompts from these problems and sample generations from five representative Code LLMs, totaling 2000 generations. 
Then, we have three experienced programmers to conduct a human evaluation to verify and analyze the excess generation issue. The result of human evaluation shows that \textbf{excess generation is a prevalent issue in contemporary code LLMs}. 
To further study the implications of excess generation issue on the inference efficiency of Code LLMs, we conduct comparative experiments. We terminate the inference process of models early when detecting excess generation, which simulates an ideal scenario without excess generation. \Fig~\ref{fig:comparison_of_different_scenarios} shows the comparative experiment result of Code Llama series models on the MBPP dataset. By comparing the inference time under the ideal scenario and the real scenario, we can find that the excess generation issue significantly increases the inference time.
This result shows that \textbf{ excess token generation is a significant bottleneck that limits the inference speed of Code LLMs in code generation tasks.}

Based on the findings in our preliminary study, we propose a straightforward and effective inference acceleration approach for Code LLMs, called \textbf{\Our}. 
{The key idea of our approach is to promptly terminate the inference when the continual generation of excess tokens is detected.} To achieve this, we design a core component \textbf{\GenGuard}, which is a unified and lightweight gating classifier attached to a Code LLM. When generating the next token, \GenGuard receives the last hidden states generated from a Code LLM and predicts whether to terminate inference at this step. To train \GenGuard, we propose an automatic training framework that includes three stages: sampling, labeling, and training. 
In the sampling stage, the framework leverages sampling prompts and a Code LLM to generate examples with excess generation issues. In the labeling stage, the framework utilizes two code analyzers to distinguish between expected and excess code in the sampled outputs and then labels them accordingly. Finally, in the training stage, we freeze the parameters of the Code LLM and train \GenGuard to predict the probability of excess tokens. This framework only necessitates users to provide sampling prompts, relieving the need and burden of manual data labeling. In the inference stage of \Our, we propose a line-wise voting mechanism that can reduce misjudgments and achieve a trade-off between generation accuracy and speed. Specifically, instead of stopping the inference process of the Code LLM immediately when \GenGuard predicts to stop inference for a single token, the line-wise voting mechanism collects all the predictions made by \GenGuard for the whole code line. After completing a code line, it uses majority voting to decide whether to terminate the inference process.

To evaluate the effectiveness of \Our, we conduct extensive experiments on five mainstream Code LLMs: the Code Llama series (7B, 13B, 34B)~\cite{roziere2023code}, \StarCoder~\cite{li2023starcoder}, and \PhindCL~\cite{phind2023codellama}, across four code generation datasets in different programming languages (PLs): MBPP~\cite{austin2021program}, MBJSP~\cite{athiwaratkun2022multi}, MBGP~\cite{athiwaratkun2022multi}, and MBCPP~\cite{athiwaratkun2022multi}. Then, we conduct ablation studies to investigate the effectiveness of each component of our approach and explore the impact of different parameter settings on the performance of our approach. Finally, we evaluate the generalizability of our approach on three untrained code generation datasets: HumanEval~\cite{chen2021evaluating}, HumanEval-JavaScript~\cite{athiwaratkun2022multi}, and HumanEval-Go~\cite{athiwaratkun2022multi}. 
The experimental results demonstrate that: (1) \Our can significantly increase the generation speed of Code LLMs without compromising the quality of the generated code. Besides, our approach is effective across different Code LLMs and different programming languages. (2) Each component of our approach plays an important role in model performance.
(3) Our approach exhibits great stability across different parameter settings. (4) Our approach exhibits great generalization capabilities on untrained datasets. 

We summarize the contributions of this paper as follows:
\vspace{-4mm}
\begin{itemize}[leftmargin=*]

\item Though extensive experiments and human evaluation in a preliminary study, we identify a significant issue of excess token generation in Code LLMs. We find that this is a prevalent issue and has a severe impact on the efficiency of code generation. To the best of our knowledge, we are the first to investigate the excess generation issue in Code LLMs.

\item We introduce \GenGuard, a unified and lightweight gating classifier designed to predict the probability of stopping generation during the generation process. To obtain training data for \GenGuard, we build an automated data construction framework. 
To make \GenGuard unified across various PLs, we curate the data from datasets in multiple PLs and validate its generality 

by comparing it with its Mono-PL counterparts.

\item We propose an effective inference acceleration method \Our by utilizing \GenGuard to promptly terminate the generation process. In addition, we adopt a line-voting mechanism that can reduce misjudgments and achieve a trade-off between generation accuracy and speed.

\end{itemize}

\secmargin
\section{Background}\label{sec:background}

\subsection{Code Large Language Models}
Code Large Language Models (Code LLMs) refer to large language models that are specifically trained for coding tasks~\cite{zhang2023survey,zheng2023survey}. Typically, Code LLM is a Transformer-based model with a large number of parameters and has strong abilities in code generation and understanding, achieved through training on a large amount of code data~\cite{zheng2023survey}. For example,  
open-source Code LLMs such as Code Llama~\cite{roziere2023code} and StarCoder~\cite{li2023starcoder}  have also shown impressive capabilities in code generation tasks. \PhindCL~\cite{phind2023codellama},  fine-tuned from \CLThirtyFour using high-quality code data, has surpassed the performance of GPT-4~\cite{openai2023gpt4} on HumanEval. 
Based on different Transformer architectures, current Code LLMs can be classified into two types: decoder-only Transformer models and encoder-decoder Transformer models~\cite{zheng2023survey}. Considering that most current Code LLMs are decoder-only Transformer models, such as Code Llama~\cite{roziere2023code} and StarCoder~\cite{li2023starcoder}, in this paper, we will focus on discussing Code LLMs with decoder-only architecture.

\secmargin
\subsection{Inference of Code LLMs}
\label{sec:background_inference}
The decoder-only Transformer architecture is initially introduced in GPT~\cite{radford2018improving}. In contrast to the original Transformer~\cite{vaswani2017attention}, the decoder-only architecture is composed solely of stacked decoder layers and does not include the encoder component~\cite{radford2018improving}. The core of the decoder layers is its masked attention layer. This layer employs masks to allow the model to focus only on previously generated tokens and prevents it from attending to future tokens during the decoding process~\cite{zhao2023survey}. This mechanism determines the autoregressive generation characteristics of the decoder-only Transformer model. 

Given an input token sequence $X = \{x_1, x_2, ..., x_n\}$, the model first maps them to context vectors $C = \{c_1, c_2, ..., c_n\}$ in word embedding layer, where $n$ is the length of the sequence. We use the word vectors as the initial hidden states $H^0 = \{ h_1^0,h_2^0,...,h_n^0\} = C$, which serve as the input to the decoder layer. And then we take the hidden states output of the previous decoder layer as input to calculate the hidden states for the next decoder layer, following the Equation~\ref{decoder_layer}, where $L$ represents the number of decoder layers, $i$ represents the current index of decoder layer and M represents the mask matrix.
\vspace{-1mm}
\begin{equation}
\small
H^i = DeocderLayer(H^{i-1},M), i \in [1, L] \label{decoder_layer}
\end{equation}
After computing the hidden states of the final decoder layer, we choose the final hidden state of the last token $h_n^L$ as the representation of the input sequence. Then, we input the representation into Language Modeling Head (LM Head) to predict the next token, following Equation\ref{lm_head}, where $W^T_{lm}$ is the LM Head Matrix. 
\begin{equation}
\small
P(x_{n+1}) = Softmax(h_n^L W^T_{lm}), \label{lm_head}
\end{equation}
After obtaining probability vectors of next token $P(x_{n+1})$, 
we decode the next token $x_{n+1}$ using greedy decoding, which means selecting the token with the highest probability following Equation~\ref{greedy_decode}.
\begin{equation}
\small
x_{n+1} = argmax(P(x_{n+1})) \label{greedy_decode}
\end{equation}

We update the output sequence $O =\{o_1\} = \{x_{n+1}\}$, and input token sequence $X = \{x_1, x_2, ..., x_n,x_{n+1}\}$, and repeat the above inference process until the inference termination condition is reached. Here are two types of inference termination conditions:
\textbf{(1) Special end token is generated}. When the special token <EOS> is generated, the model will stop generation. \textbf{(2) The length of the generated tokens exceeds the maximum value}. For example, in \textit{Transformer}\footnote{\url{https://github.com/huggingface/transformers}} library, there is a parameter \textit{max\_new\_tokens}, when the length of the generated token sequence exceeds the value of \textit{max\_new\_tokens}, the inference of the model will be terminated.

\secmargin
\section{Experimental Design}
\label{sec:experiments_settings}

\subsection{Datasets}

We use four public datasets on code generation, including MBPP~\cite{austin2021program} in Python, MBJSP~\cite{athiwaratkun2022multi} in JavaScript, MBGP~\cite{athiwaratkun2022multi} in Go, and MBCPP~\cite{athiwaratkun2022multi} in C++. We use the above datasets in both the preliminary study and evaluation sections. Additionally, to further validate the generalizability of our proposed approach, we utilize HumanEval~\cite{chen2021evaluating}, HumanEval-Go~\cite{athiwaratkun2022multi} and HumanEval-JavaScript~\cite{athiwaratkun2022multi}. The statistics of  datasets are listed in Table~\ref{table:benchmark_statistics}, and the detailed description of these datasets are as follows:

\begin{itemize}[leftmargin=*]
    \item \textbf{MBPP}~\cite{austin2021program} is a frequently used benchmark in code generation, comprising 974 crowd-sourced Python programming problems. Each problem includes a task description, a solution, and three test cases. 
    
    \item \textbf{HumanEval}~\cite{chen2018tree} is a popular benchmark in code generation, which includes 164 hand-written Python programming problems. Every problem contains a function signature, docstring, function body, and several unit tests.

    \item \textbf{MBXP and Multilingual HumanEval} ~\cite{athiwaratkun2022multi} are the multilingual versions of the MBPP and HumanEval benchmarks, sharing a similar data format. For MBXP, we use \textbf{MBJSP}, \textbf{MBGP}, and \textbf{MBCPP}, which are datasets in three popular programming languages. Because these three datasets are not pre-partitioned, we selected 500 samples to form the test set. The remaining samples are split in a 9:1 ratio to create the training and validation sets, respectively. As for Multilingual HumanEval, due to the lack of a C++ dataset, we use \textbf{HumanEval-JavaScript} and \textbf{HumanEval-Go} to verify the PL generalization ability of our approach.
\end{itemize}

\begin{table}[t]
\small
\centering
\caption{Dataset statistics.}
\label{table:benchmark_statistics}
\tabmargin
\begin{tabular}{lclccc}
\toprule
Benchmark  & Language & \#Train & \#Valid & \#Test \\ \midrule
MBPP  & Python & 374 & 90 & 500 \\ 
MBJSP  & JavaScript & 420 & 46 & 500 \\ 
MBGP  & Go & 396 & 43 & 500 \\ 
MBCPP   & C++ & 314 & 34 & 500 \\ 
HumanEval  & Python & \textbackslash & \textbackslash & 164 \\ 
HumanEval-JS & JavaScript & \textbackslash & \textbackslash & 161 \\ 
HumanEval-Go  & Go & \textbackslash & \textbackslash & 160 \\ 
\bottomrule
\end{tabular}

\end{table}
\secmargin
\subsection{Metrics and Notations}\label{sec:metrics}
Following previous code generation studies~\cite{li2023structured,li2023large}, we employ \textbf{Pass@k} (k = 1, 3, 5) as our evaluation metric to measure code generation accuracy. A requirement is considered solved if the generated program passes all test cases, and Pass@k is the percentage of solved problems in total problems. 
Additionally, we use the following metrics/notations to measure code generation quality and speed. \textbf{Length} represents the average number of tokens of generated solutions. \textbf{Time} represents the average time (in seconds) taken by a Code LLM to generate a program per sample. \textbf{Speedup} measures the improvement in code generation speed of a Code LLM, relative to its corresponding baseline performance. \textbf{ER (Excess Ratio)} quantifies the percentage of generated results that contain excess tokens. \textbf{PGWE} (percentage of generation without <EOS> Ending) indicates the percentage of results where an excess generation issue is detected and the special ending token <EOS> is not generated. This means the model will keep generating excess code until the token length reaches the maximum generation length.

\vspace{-3mm}
\subsection{Base LLMs for Code}\label{sec:base_code_llms}
In the preliminary study and evaluation, we select five mainstream Code LLMs as the base models for our experiments. The details of these models are as follows:

\begin{itemize}[leftmargin=*]
    \item \textbf{Code Llama Series}~\cite{roziere2023code} is a series of models pre-trained on code data using Llama 2~\cite{touvron2023llama}, with parameter sizes ranging from 7B, 13B, to 34B. These models are currently the most powerful open-source foundation Code LLMs available. In our experiments, we select the multilingual versions of the models 
    \CLSeven, \CLThirteen, and \CLThirtyFour.

    \item \textbf{StarCoder}~\cite{li2023starcoder} is a  15.5B parameter model trained on 80+ programming languages, which has demonstrated remarkable performance on code generation tasks. In our experiments, we select \StarCoder as a base model.
    
    \item \textbf{\PhindCL}~\cite{phind2023codellama} is an advanced Code LLM fine-tuned from \CLThirtyFour on a proprietary dataset of about 80k high quality programming problems and solutions. This model outperforms GPT-4 on the HumanEval benchmark, achieving state-of-the-art performance. 
    In our experiments, we use \PhindCL-v2 as the base model, which is the state-of-the-art version of the Phind series of models.
\end{itemize}

Although ChatGPT~\cite{openai2022chatgpt} and GPT-4~\cite{openai2023gpt4} have achieved notable performance in code generation tasks, we do not use them in our experiments. This is because OpenAI only provides API access for these models and does not open-source their parameters. This limitation restricts our exploration of the details of these models' inference processes. 

\vspace{-2mm}
\subsection{Experimental Settings}
\label{sec:settings}
We provide detailed experimental settings as follows.

\textbf{Code generation settings.} In our experiments, we deploy Code LLMs on the GPU using the bfloat16~\cite{googlecloud2020bfloat16} precision format. The max generated length is set to 300. We employ greedy decoding for calculating Pass@1 and nucleus sampling for calculating Pass@3 and Pass@5. 
Empirically, we set the temperatures following the original settings, which is 0.8 for function-level tasks~\cite{li2023large,li2023structured} and 0.2 for class-level tasks~\cite{du2023classeval}. Notably, following previous studies~\cite{li2023towards, athiwaratkun2022multi,li2023structured,zhu2023improving}, our prompts for function-level code generation is designed in a partial code format, including the signature of the target function, comments with requirements, and relevant test cases. Across the experiments conducted in \Sec~\ref{sec:effectiness_experiments},~\ref{sec:ablation_study_experiments}, and~\ref{sec:stability_experiments}, we set the parameter $\theta_{stop}$ to 0.5. In experiments of \Sec~\ref{sec:experiments:genralizability}, we set a $\theta_{stop}$ of 0.7 when evaluating our approach on untrained datasets. 

\textbf{Training stage settings.} In the sampling stage, we sample raw generations for each Code LLM mentioned in \Sec~\ref{sec:base_code_llms} from the training sets of MBPP, MBJSP, MBGP, and MBCPP. Subsequently, each Code LLM's raw generations from different programming languages are merged and labeled to construct a dataset, referred to as the Multi-PL training data. When labeling data with ChatGPT, we choose \textit{gpt-3.5-turbo-16k}~\cite{openai2022gpt3.5turbo} 
to label data. 
In the training stage, the parameters of each Code LLM are frozen, and a linear classifier is trained to act as its \GenGuard. 
The training process uses the AdamW~\cite{loshchilov2017decoupled} optimizer with a learning rate of 5e-4 and lasts for ten epochs. Our experiments are conducted on a machine with 216 GB main memory and a Tesla A100 80GB GPU.

\begin{figure*}[t]
    \centering
    \includegraphics[
    width=0.95\linewidth]{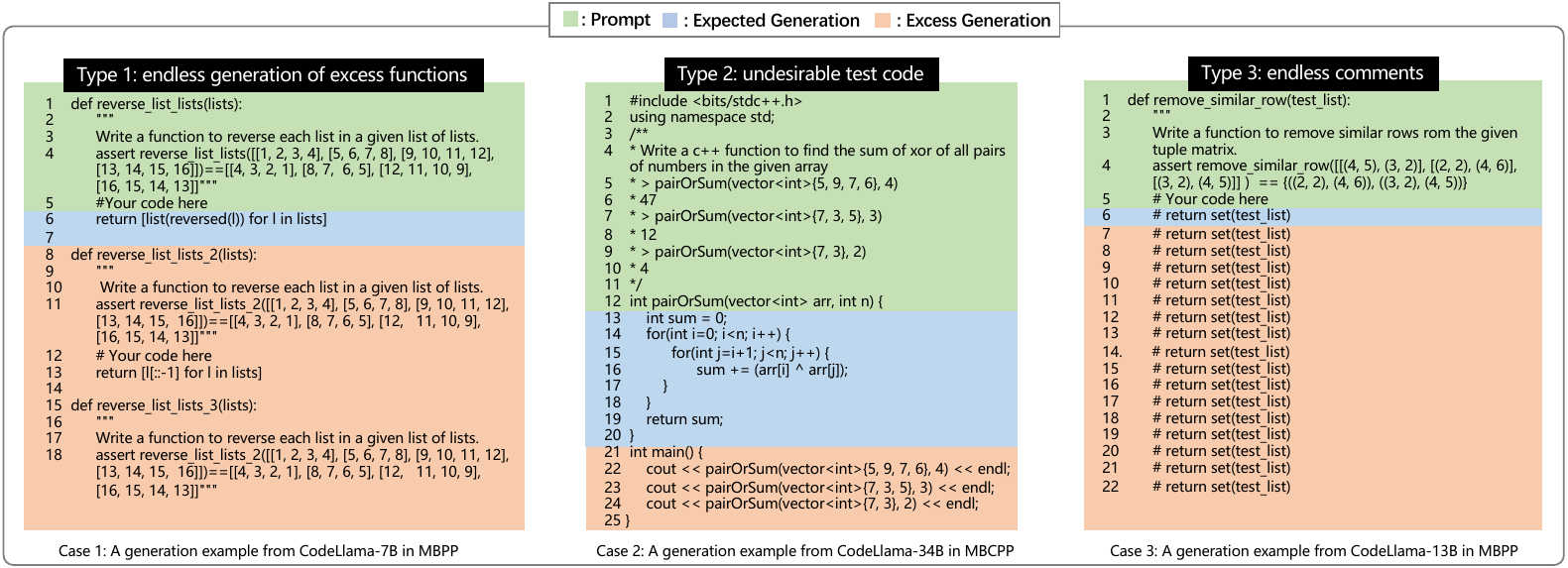}
    \figmargin
   
    \caption{Results of qualitative analysis in the preliminary study.}
    \label{fig:pre_study_examples_overview}
    \figmargin
\end{figure*}

\section{Preliminary Study}\label{sec:preliminary}

In this section, we find an efficiency issue in the inference of Code LLMs, which is the continual generation of excess tokens. This issue means that after the model generates code corresponding to a given requirement, the generation of tokens does not cease. Instead, it continues to produce undesired content that offers no assistance in solving the current problem. Considering the high inference time complexity of Code LLMs, this issue significantly increases inference time. 

To gain a deeper and more comprehensive understanding of this efficiency issue in Code LLMs, we conduct extensive preliminary studies. First, we randomly select 100 
problems from training sets of MBPP, MBJSP, MBGP, and MBCPP. Next, we build prompts from these problems and sample generations from the five Code LLMs described in \Sec~\ref{sec:base_code_llms}, totaling 2000 generations. Then, we involve three annotators, each with over five years of coding experience, and manually identify and label the excess code tokens in generations. Annotators consist of two authors and a Ph.D. student in computer science. Given prompts and generated code, annotators independently truncate the generated code to remove excess tokens without changing its functionality and annotate truncated line numbers as labels. For any inconsistent or uncertain annotations, we conduct face-to-face discussions to resolve disagreements.
The agreement among annotators is verified using Krippendorff's alpha~\cite{hayes2007answering}, and the value is 0.998, indicating a high agreement among annotators and confirming the reliability of annotated results. 

\begin{itemize}[leftmargin=*]
    \item \textbf{PSRQ1:} Is the phenomenon of excess token generation a prevalent issue in contemporary code LLMs?
    \item \textbf{PSRQ2:} What are the implications of excess token generation on the inference efficiency of Code LLMs?

\end{itemize}

\vspace{-2mm}
\subsection{Excess Token Phenomenon (PSRQ1)}
\subsubsection{Qualitative Analysis}\label{sec:case_study}

Through case studies, we find that there are three major types of issues.
\textbf{Type 1: endless generation of excess functions.} 
The first case in \Fig~\ref{fig:pre_study_examples_overview}  presents a typical example, showcasing a result generated by \CLSeven on the MBPP dataset. 
After generating the expected function body (line 6), the generation process does not stop. The model continues to generate tokens until line 15, where it is forcibly truncated when the maximum sequence length is reached. 
It is notable that the extra content generated in lines 7-15 are excess functions with unhelpful content. 
Considering the high inference time complexity of Code LLM, such redundant generation significantly increases the inference time of models.
\textbf{Type 2: undesirable test code.}  The second case in \Fig~\ref{fig:pre_study_examples_overview} presents a typical example generated by Phind CodeLlama-34B on MBCPP. After generating the expected function (lines 12-19), the model continues to produce test code (lines 20-24). Although these test codes are somewhat related to expected generations, they are generally useless and lead to additional inference time.
\textbf{Type 3: endless comments.} 
The third case in \Fig~\ref{fig:pre_study_examples_overview} presents a typical example generated by \CLThirteen on MBPP. After receiving the prompt (lines 1-5), the model generates an extensive amount of repetitive comments (lines 6-15), continuing until the maximum sequence length is reached. In this case, the model fails to generate code that solves the problem and instead produces endless comments. Ideally, we would want the model to stop inference before generating repetitive comments, thereby reducing unnecessary time consumption.

We believe that this issue mainly stems from the pre-training process of Code LLMs. During its pre-training, Code LLMs typically utilize a large dataset of code files~\cite{li2023starcoder,touvron2023llama}, which often contain multiple functions, test codes, and comments. As a result, when these models are used for function-level code generation, they tend to produce content akin to training data, leading to the generation of excess functions, test codes, etc.

\subsubsection{Quantitative Analysis}
To investigate to what extent this issue exists in different Code LLMs across various programming languages, we randomly select 100 prompts from MBPP, MBJSP, MBGP, MBCPP, and sample generations from five Code LLMs as shown in \Sec~\ref{sec:base_code_llms}, totaling 2000 generations. Then, we manually check the outputs and mark the expected generation and excess generation for each sample. We can derive two observations from the human evaluation results shown in Table~\ref{table:pre_study}.

\textbf{Observation 1: all Code LLMs have the excess generation issue across different programming languages}. From Table~\ref{table:pre_study}, we find that all Code LLMs exhibit a high ER (from 59\% to 100\%), indicating a high frequency of excess generation. 
For example, outputs of \CLSeven in MBPP show an ER of 100\%, which means all results include excess generation. Moreover, we find that models with smaller parameter sizes, such as \CLSeven, \CLThirteen, and \StarCoder, exhibit a higher PGWE. This means that, in some extreme cases, these models not only output excess generations but also continue generating content until reaching the maximum length of generated tokens. 

\textbf{Observation 2: supervised fine-tuning can mitigate the issue of excess generations in Code LLMs}. \PhindCL is a model fine-tuned on \CLThirtyFour using a proprietary dataset of high-quality programming problems and solutions. From Table~\ref{table:pre_study}, we can see that \PhindCL, compared to \CLThirtyFour, exhibits a lower ER and shorter token length on the MBPP and MBJSP datasets. Additionally, it also demonstrates a lower PGWE on both MBPP and MBJSP. We believe this is due to the high-quality fine-tuning data enabling the model to better understand the intent of requirements, thereby mitigating the issue of excess generations.

\eqmargin
\begin{center}
    \begin{myboxc}\textbf{PSRQ1 Summary: }
    Overall, all five Code LLMs have severe issues of 
    excess generations across four programming languages.
    \end{myboxc}
\end{center}
\eqmargin

\begin{table*}[t]
\small \setlength{\tabcolsep}{2pt}
\caption{Human evaluation results of preliminary study. ``Original'' represents the raw output result of the model; ``Expected'' represents the result after manually removing excess tokens. }
\label{table:pre_study}
\tabmargin
\resizebox{\linewidth}{!}{
\begin{tabular}{l c cccc cccc cccc cccc}\toprule
  \multirow{2}{*}{\textbf{Model}} & \multirow{2}{*}{\textbf{Type}} 
        & \multicolumn{4}{c}{\textbf{MBPP (Python)}} & \multicolumn{4}{c}{\textbf{MBJSP (JavaScript)}} 
        & \multicolumn{4}{c}{\textbf{MBGP (Go)}} & \multicolumn{4}{c}{\textbf{MBCPP (C++)}} \\  
  \cmidrule(r){3-6} \cmidrule(r){7-10} \cmidrule(r){11-14} \cmidrule(r){15-18} 
 & & \textbf{Length} & \textbf{Time} & \textbf{ER} & \textbf{PGWE}  
   & \textbf{Length} & \textbf{Time} & \textbf{ER} & \textbf{PGWE}  
   & \textbf{Length} & \textbf{Time} & \textbf{ER} & \textbf{PGWE}  
   & \textbf{Length} & \textbf{Time} & \textbf{ER} & \textbf{PGWE} \\ \midrule
\multirow{2}{*}{\CLSeven} 
  & Original  & 300 & 9.31 & 100\% & 100\% & 276.8 & 8.49 & 97\% & 81\% 
              & 270.7 & 7.96 & 98\% & 81\% & 292.8 & 8.75 & 100\% & 96\% \\
  & Expected  & 51.4 & 1.59 & 0\% & 0\% & 87.6 & 2.65 & 0\% & 0\% 
              & 72.2 & 2.21 & 0\% & 0\% & 82.5 & 2.56 & 0\% & 0\% \\\hline
\multirow{2}{*}{\CLThirteen}
  & Original  & 300 & 12.15 & 100\% & 100\% & 300 & 11.92 & 100\% & 100\%
              & 297.2 & 10.87 & 99\% & 97\% & 298.3 & 11.51 & 100\% & 99\% \\ 
  & Expected  & 41.8 & 1.65 & 0\% & 0\% & 83.8 & 3.29 & 0\% & 0\%
              & 78.7 & 3.03 & 0\% & 0\% & 79.4 & 3.15 & 0\% & 0\% \\\hline
\multirow{2}{*}{\CLThirtyFour} 
  & Original  & 269.8 & 14.97 & 97\% & 82\% & 158.2 & 8.87 & 98\% & 3\% 
              & 260.1 & 14.58 & 98\% & 61\% & 180.4 & 10.08 & 100\% & 14\% \\
  & Expected  & 52.0 & 2.89 & 0\% & 0\% & 81.8 & 4.60 & 0\% & 0\%
              & 81.0 & 4.60 & 0\% & 0\% & 72.8 & 4.12 & 0\% & 0\%  \\\hline
\multirow{2}{*}{\StarCoder} 
  & Original  & 202.3 & 5.29 & 98\% & 53\% & 231.8 & 6.23 & 100\% & 52\%
              & 284.6 & 7.69 & 98\% & 87\% & 257.9 & 7.19 & 100\% & 68\% \\
  & Expected  & 81.3 & 2.32 & 0\% & 0\% & 78.3 & 2.38 & 0\% & 0\% 
              & 106.5 & 2.90 & 0\% & 0\% & 108.2 & 3.42 & 0\% & 0\% \\\hline
\multirow{2}{*}{\PhindCL} 
  & Original  & 120.7 & 6.70 & 59\% & 4\% & 184.2 & 10.42 & 81\% & 7\% 
              & 245.0 & 13.71 & 99\% & 45\% & 187.6 & 10.52 & 90\% & 13\% \\
  & Expected  & 60.7 & 3.39 & 0\% & 0\% & 93.7 & 5.28 & 0\% & 0\%
              & 81.3 & 4.62 & 0\% & 0\% & 74.8 & 4.28 & 0\% & 0\% \\ \bottomrule
\end{tabular}
}
\end{table*}
\vspace{-2mm}

\subsection{Implications on Inference Efficiency (PSRQ2)}

Firstly, to explore the impact of this issue on the length of the generated token sequences, we calculate and compare the average lengths of outputs with and without excess generations. Additionally, to investigate the impact on the inference speed of Code LLMs, we prematurely terminated the model's output generation upon the occurrence of excess generations and measured its effect on inference speed. From the experimental results shown in Table~\ref{table:pre_study}, we have  \textbf{Observation 3: the issue of excess generations significantly increases the length of generated sequences and hence significantly increases generation time.} For instance, the average output sequence length for \CLThirtyFour on MBJSP is 158.2 (``Original''). Ideally, if there were no excess generation, the output length should decrease to 81.8 (``Expected''). Considering the high inference time complexity of Code LLMs, these additional generated tokens will significantly increase the inference time of Code LLMs. By prematurely terminating the inference process upon the occurrence of excess generations, all Code LLMs obtain a significant increase in inference speed. For example, through early stopping, the inference speed of \CLThirtyFour on MBGP increased to 3.17 times its original speed.

\eqmargin
\begin{center}
    \begin{myboxc}\textbf{PSRQ2 Summary: }
    The issue of excess generations is a primary factor limiting the generation efficiency of Code LLMs.
    \end{myboxc}
\end{center}
\eqmargin

\section{Approach}\label{sec:approach}
\begin{figure*}[t]
    \centering
    \includegraphics[width=0.9\linewidth]{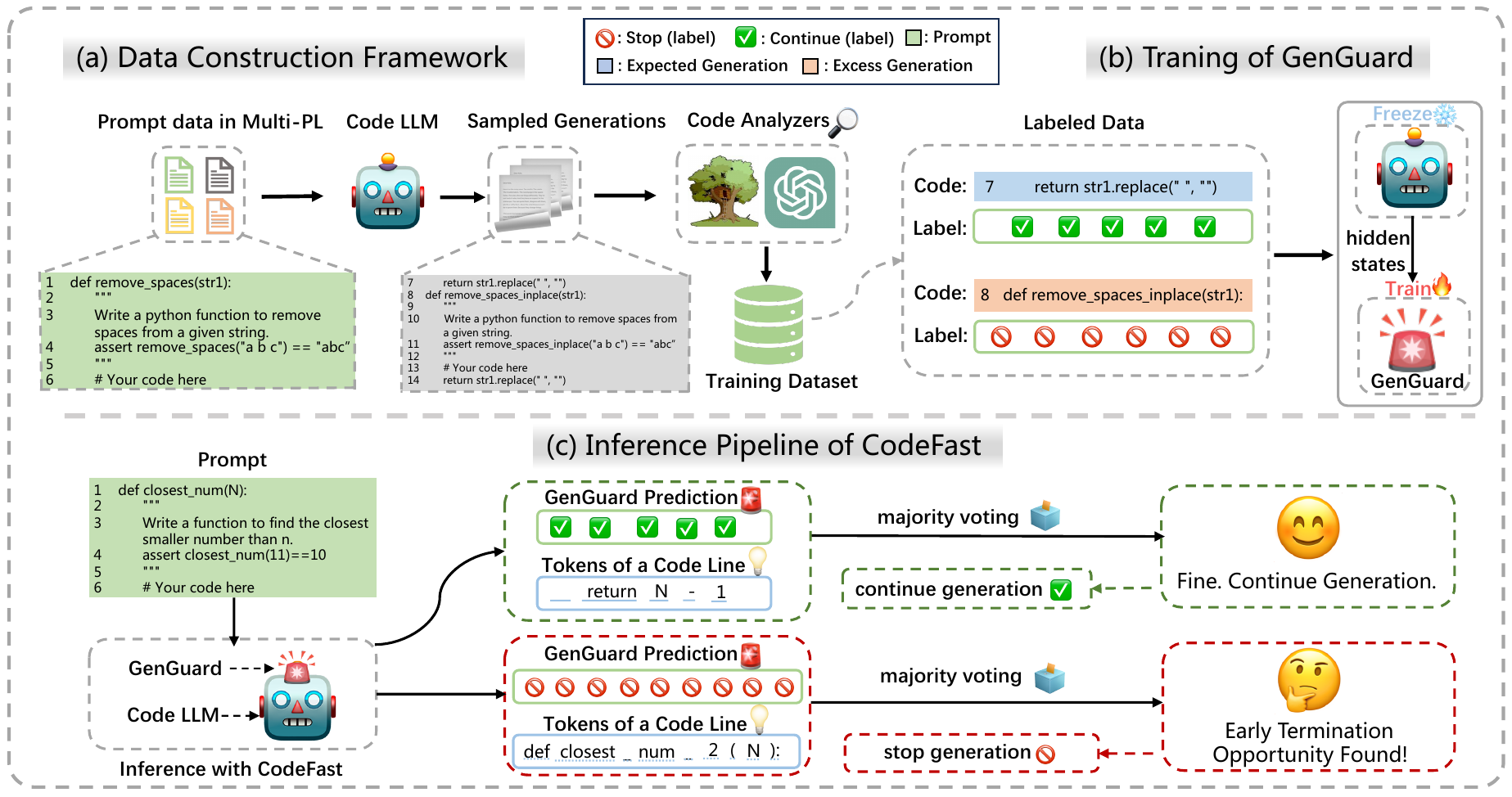}
    \figmargin
    \caption{Overview of \Our.}
    \label{fig:overview}
\end{figure*}

In this section, we propose an effective code generation acceleration approach called \textbf{\Our} as shown in \Fig~\ref{fig:overview}. The key idea is to terminate the inference early when a continual
generation of excess tokens is detected. \Our achieves this through a key component: \GenGuard, which is an additional module to predict whether to stop inference at the current step. 

\vspace{-2mm}

\subsection{Data Construction Framework}
\label{sec:data_construction}
Our data construction framework is shown in \Fig~\ref{fig:overview}(a). By simply providing sampling prompts, our approach can automatically generate labeled data to train a specialized \GenGuard for a Code LLM. 
This approach includes two stages, \textit{the sampling stage} and  \textit{the labeling stage}. 

\subsubsection{The sampling stage}
In the sampling stage, we utilize sampling prompts to collect 
raw generations from the model, which potentially has the issue of excess generations. The sampling prompts are in the format of partial code snippets, each including a function signature and comments that contain requirements and test cases. We collect sampling prompts from the training sets of MBPP, MBJSP, MBGP, and MBCPP. Then, we input these prompts into Code LLMs to obtain a large number of raw outputs. 

\subsubsection{The labeling stage}
A key challenge for labeling data is how to differentiate between expected generation and excess generations in raw outputs. We design a labeling pipeline as shown in Algorithm~\ref{algorithm:labeling}. We utilize two types of code analyzers to label raw outputs from Code LLMs. \textbf{(1) Syntax-based code analyzer.} It can label the first two types of generations shown in \Sec~\ref{sec:case_study} easily by analyzing the syntax structure of generated code. This code analyzer considers all code lines related to the target function as expected generation. This includes not only the main body of the target function but also all functions that it calls. The rest of code lines are then labeled as excess generations. To achieve this, we utilize tree-sitter\footnote{\url{https://github.com/tree-sitter/tree-sitter}}, an efficient parser generator that can analyze the structure of code via abstract syntax trees (ASTs). First, it builds the ASTs from the raw outputs and extracts the node of the target function using its signature. Then, it extracts all nodes called by the target function by traversing the AST. Next, it collects all extracted function nodes and identifies the last line of these functions. Finally, the portion from the beginning up to the end of this specific line is labeled as the expected generation, and the remaining part of the output is labeled as excess generation. \textbf{(2) Semantic-based code analyzer.} It is utilized when the first analyzer fails. For instance, the third type of generation, as illustrated in Fig~\ref{fig:pre_study_examples_overview}, is challenging to label by relying solely on the syntax structure of the code. It requires an understanding of the semantics of the generated code. Here, we utilize ChatGPT~\cite{openai2022chatgpt}, a powerful LLM with great code understanding ability, to label the data. Inspired by the in-context-learning technique~\cite{xiao2023freeal}, we teach ChatGPT how to label the data with detailed task instructions and manually crafted demonstrations. The concrete prompt utilized in ChatGPT can be found in the Appendix of replication package~\cite{CodeFast}. Additionally, we conduct a human evaluation to verify the quality of the results truncated by ChatGPT. These annotators include two authors and one Ph.D. student in computer science. Given raw code and ChatGPT's truncated code,  they are tasked with assessing the accuracy of the truncations. Each truncation is scored on a binary scale: 1 for correct and 0 for incorrect. The final average score is 0.957, indicating a high quality of ChatGPT's labeling. Moreover, we calculate the value of Krippendorff’s alpha~\cite{hayes2007answering} and obtain a result of 0.919. This indicates a high level of agreement among the annotators and confirms the reliability of the results truncated by ChatGPT.

Finally, we label each token of the expected generation portion as ``continue generation'' and the excess generation portion as ``stopping generation''. Note that we only keep the first excess line since we expect \GenGuard to stop at the earliest step. 

\begin{algorithm}
\small
\caption{Labeling Algorithm}
\label{algorithm:labeling}
\begin{algorithmic}[1] 
\Require Raw Output: $o_{raw}$, Partial Code Prompt: $prompt_{code}$, AST Builder: $ast\_builder$, ChatGPT: $chatgpt$, Prompt for ChatGPT: $prompt_{gpt}$
\Ensure Labeled data: $d_{label}$
\State $output\_ast \gets ast\_builder(output)$
\State $signature \gets extract\_signature(prompt_{code})$
\State $target\_func\_node \gets output\_ast.locate\_node(signature)$
\State $o_{expected} \gets None$, $o_{excess} \gets None$
\State $last\_line\_index \gets -1$, $node\_stack \gets [target\_func\_node]$
\While{$node\_stack \neq []$}
    \For{$node$ in $node\_stack$}
        \State $last\_line\_idx \gets \max(last\_line\_idx, node.last\_line)$
        \State $called\_node\_list .append( node.find\_called\_nodes()$)
        \State Delete  $node$ from $node\_stack$
        \State $node\_stack$ += $called\_node\_list$
    \EndFor
\EndWhile
\State $o_{expected},o_{excess} \gets split(o_{raw},last\_line\_idx)$
\If{$o_{expected} = prompt_{code}$}
    \State $prompt \gets prompt_{gpt}.replace('\textless{}raw\_output\textgreater{}',o_{raw})$
    \State $o_{expected},o_{excess} \gets chatgpt(prompt)$
\EndIf
\If{$o_{expected} \neq prompt_{code}$}
    \State $o_{excess} \gets extract\_first\_line(o_{excess})$
    \State $d_{label} \gets \{\text{``continue''}: o_{expected}, \text{``stop''}: o_{excess}\}$
\Else
    \State $d_{label} \gets \{\text{``continue''}: None, \text{``stop''}: None\}$ 
\EndIf
\State \Return $d_{label}$
\end{algorithmic}
\end{algorithm}

\subsection{Training of \GenGuard Module}
\label{sec:training_framework}

The key component of \Our is the \GenGuard Module, which is an additional module attached to Code LLM. 
The \GenGuard functions as a lightweight gating classifier to control the inference process of Code LLM. To reduce extra parameters, we train a straightforward linear classifier for each Code LLM as its specialized \GenGuard Module. Following previous studies~\cite{hao2023toolkengpt}, we freeze the parameters of Code LLM and only train the \GenGuard. In this process, Code LLM is viewed as a feature extractor and encodes the code sequence into a feature vector, which represents the semantics of this code. Subsequently, the \GenGuard module is trained using these feature vectors as input. The details of \GenGuard training are as follows:

Given a code sequence $X_{n}=(x_{1},x_{2},...x_{n})$ and label $y_{n}$  at step $n$, we input this sequence into 
$LLM$ to obtain the last hidden state $h^{L}_{n}$, following Equation~\ref{equa:hidden_state_training}, where $L$ stands for the number of decoder layers in Code LLMs.
\begin{equation}
\label{equa:hidden_state_training}
\small
    h^{L}_{n} = LLM(X_{n})
\end{equation}
Then, we input the last hidden state into the linear classifier to predict the probability of stopping $p_{stop}$ at this step: 
\begin{equation}
\label{equation:prob_predict_training}
\small
    P_{stop} = \{p_{continue}, p_{stop}\}= Softmax(h_{n}^{L}W_{GenGuard}^T)
\end{equation}
Finally, we use the cross entropy loss 
to optimize the parameters of linear classifiers, the loss is computed following~\ref{equation:loss}, where $y_{n}$ equals 1 to indicate ``stop generation'' and 0 to indicate ``continue generation''.
\begin{equation}
\label{equation:loss}
\small
    Loss = - ( y_{n} log(p_{stop})+(1-y_{n})log(1-p_{stop}) )
\end{equation}

The statistics show that the largest GenGuard module has fewer than 0.02M parameters, amounting to just 0.00005\% to 0.0001\% of the parameter size of our studied Code LLMs. This indicates that \GenGuard is lightweight\footnote{We present the detailed parameter information of GenGuard modules attached to five Code LLMs in Appendix-C of replication package~\cite{CodeFast}.}.

\subsection{Inference Pipeline of \Our}
\label{sec:approach_in_inference}
In this section, we introduce the code generation process 
of \Our. 
First, given an input sequence $X_n = \{x_1, x_2, ... x_n\}$ at step $n$, we input this sequence into Code LLM to predict next token $x_{n+1}$ and last hidden states $h_{n}^{L}$, following Equation~\ref{equation:hidden_states_output}, where $L$ represents the layer numbers of Code LLM.
\begin{equation}
\label{equation:hidden_states_output}
\small
    x_{n+1}, h_{n}^{L} = LLM(X_n)
\end{equation}
The last hidden states $h_{n}^{L}$ at step $n$ is a high dimensional feature vector, which includes extensive semantic information of the current input sequence $X_n$. Following \cite{hao2023toolkengpt}, we input the feature vectors extracted from LLM into \GenGuard $W_{GenGuard}$, to predict the probability of stopping $p_{stop}$ at the current step $n$: 
\begin{equation}
\label{equation:stop_prob_prediction}
\small
    P_{stop} = \{p_{continue}, p_{stop}\}= Softmax(h_{n}^{L}W_{GenGuard}^T)
\end{equation}
Then, we compare the probability of stopping $p_{stop}$ with hyperparameter stopping threshold $\theta_{stop}$ and terminate the inference of code generation when $p_{stop}$ exceeds $\theta_{stop}$.

However, there is a potential problem that may reduce the quality of the generated code. If \GenGuard erroneously predicts ``stop generation'' when it should predict ``continue generation'', the Code LLM will terminate the inference process too early, leading to incomplete code. Inspired by the concept of majority voting, 
we propose an algorithm named ``line-voting mechanism'' to mitigate this issue. The key idea of this method is to emulate the way programmers write code line by line. Instead of stopping the inference process of the Code LLM immediately when \GenGuard predicts ``stop generation'', 
the line-voting mechanism collects all the predictions made by \GenGuard during the generation of a single code line. After completing the code line, it uses majority voting to decide whether to terminate the inference process. Although the line-voting mechanism may lead to the generation of more excess tokens and slightly increase inference time, it reduces the likelihood of false predictions by \GenGuard, thereby enhancing the quality of the generated code.

The inference process of the \Our-enhanced Code LLM is illustrated in \Fig~\ref{fig:overview}(c). Upon receiving the prompt (lines 1-6), the model begins to generate tokens for the next line (line 7), while the \GenGuard module simultaneously predicts whether to stop at each step. When a new line is detected, we perform majority voting on all predictions. In this case, the generation process will continue when the voting result is ``continue generation'' and terminate when the voting result is ``stopping generation''.

\secmargin
\section{Evaluation}
\label{sec:evaluation}

We summarize the following research questions (RQs) to evaluate \Our:
\begin{itemize}[leftmargin=*]
    \item \textbf{RQ1:} What is the effectiveness of \Our?
    \item \textbf{RQ2:} How much do different components contribute?
    \item \textbf{RQ3:} What is the stability of \Our?
    \item \textbf{RQ4:} What is the generalizability of \Our?
    \item \textbf{RQ5:} Can \Our be applied to other scenarios?

\end{itemize}

\secmargin
\subsection{RQ1: Effectiveness}
\label{sec:effectiness_experiments}

\begin{table*}[t]
\centering \small
\setlength{\tabcolsep}{1.5pt}
\caption{Performance of \Our on five Code LLMs across four benchmarks. P@1 is short for Pass@1. Detailed metrics descriptions are provided in \Sec~\ref{sec:metrics}.}\label{table:effectiveness}
\tabmargin
\resizebox{0.95\linewidth}{!} {
\begin{tabular}{l cccc cccc cccc cccc}\toprule
  \multirow{2}{*}{\textbf{Model}} & \multicolumn{4}{c}{\textbf{MBPP (Python)}} & \multicolumn{4}{c}{\textbf{MBJSP (JavaScript)}} & \multicolumn{4}{c}{\textbf{MBGP (Go)}} & \multicolumn{4}{c}{\textbf{MBCPP (C++)}} \\  
  \cmidrule(r){2-5} \cmidrule(r){6-9} \cmidrule(r){10-13} \cmidrule(r){14-17} 
  & \textbf{P@1} & \textbf{Time} &\textbf{Speedup} & \textbf{Length} 
  & \textbf{P@1} & \textbf{Time} &\textbf{Speedup} & \textbf{Length} 
  & \textbf{P@1} & \textbf{Time} &\textbf{Speedup} & \textbf{Length} 
  & \textbf{P@1} & \textbf{Time} &\textbf{Speedup} & \textbf{Length} \\ \midrule
\CLSeven  &{41.2} &{{9.36}} &{$\times$1.00} & {299.2} 
          &{47.2} &{{8.81}} &{$\times$1.00} & {283.4} 
          & {33.4} &{{7.1}} &{$\times$1.00} & {267.0} 
          &{43.0} &{{8.44}} &{$\times$1.00} & {290.6}   \\ 
\rowcolor{gray!15}
\CLSevenOurs  &{41.0} &{2.07} &{\textbf{$\times$4.52}} & 63.5 
              &{47.2} &{3.36} &{\textbf{$\times$2.62}} & 110.7
              &{33.4} &{3.42} &{\textbf{$\times$2.07}} & 108.5 
              &{43.0} &{3.17} &{\textbf{$\times$2.66}} & {98.8} \\\hline
\CLThirteen &{44.8} &{11.87} &{$\times$1.00} & 300 
            &{{51.6}} &{11.92} &{$\times$1.00} & 300
            &{{38.4}} &{11.94} &{$\times$1.00} & 298.7 
            &{{51.4}} &{11.70} &{$\times$1.00} & 299.4 \\ 
\rowcolor{gray!15}
\CLThirteenOurs &{44.8} &{2.98} &{\textbf{$\times$3.98}} & 71.9 
                &{{51.6}} &{4.31} &{\textbf{$\times$2.77}} & 105.2 
                &{{38.4}} &{4.38} &{\textbf{$\times$2.73}} & 105.3 
                &{{51.4}} &{4.00} &{\textbf{$\times$2.93}} & 97.2 \\\hline
\CLThirtyFour &{{51.6}} &{14.67} &{$\times$1.00} & 264.8 
              &{{57.4}} &{8.37} &{$\times$1.00} & 149.2 
              &{{42.4}} &{14.51} &{$\times$1.00} & 263.4 
              &{{55.2}} &{10.12} &{$\times$1.00} & 180.6 \\
\rowcolor{gray!15}
\CLThirtyFourOurs &{{51.6}} &{3.62} &{\textbf{$\times$4.05}} & 63.7 
                  &{{57.4}} &{5.58} &{\textbf{$\times$1.50}} & 91.5 
                  &{{42.4}} &{5.89} &{\textbf{$\times$2.46}} & 97.6 
                  &{{55.2}} &{4.95} &{\textbf{$\times$2.04}} & 79.9 \\\hline
\StarCoder  &{{42.6}} &{5.14} &{$\times$1.00} & 204.3 
            &{{32.8}} &{6.42} &{$\times$1.00} & 252.8   
            &{{31.2}} &{7.17} &{$\times$1.00} & 284.6 
            &{{45.2}} &{6.15} &{$\times$1.00} & 249.9  \\ 
\rowcolor{gray!15}
\StarCoderOurs  &{{42.6}} &{1.98} &{\textbf{$\times$2.59}} & 73.5 
                &{{32.8}} &{1.56} &{\textbf{$\times$4.11}} & 58.0  
                &{{31.2}} &{2.88} &{\textbf{$\times$2.49}} & 108.4 
                &{{45.2}} &{2.55} &{\textbf{$\times$2.41}} & 93.9 \\ \hline
\PhindCL  &{{55.8}} &{4.89} &{$\times$1.00} & 88.0 
          &{{60.0}} &{8.34} &{$\times$1.00} & 147.9 
          &{{42.2}} &{14.19} &{$\times$1.00} & 250.7 
          &{{60.4}} &{10.04} &{$\times$1.00} & 178.6 \\
\rowcolor{gray!15}
\PhindCLOurs  &{{55.8}} &{3.65} &{\textbf{$\times$1.34}} & 64.3 
              &{{60.0}} &{5.79} &{\textbf{$\times$1.44}} & 96.6 
              &{{42.2}} &{5.78} &{\textbf{$\times$2.46}} & 94.8 
              &{{60.4}} &{5.12} &{\textbf{$\times$1.96}} & 83.0  \\ \bottomrule
\end{tabular}
}
\end{table*}

To answer RQ1, we evaluate the quality of code generated by Code LLMs and the generation efficiency. We evaluate the quality of generated code with Pass@k (k=1,3,5) and evaluate the generation efficiency with Length, Time, and Speedup as described in \Sec~\ref{sec:metrics}. 
Moreover, we use five popular Code LLMs shown in \Sec~\ref{sec:base_code_llms} (\CLSeven, \CLThirteen, \CLThirtyFour, \StarCoder, and \PhindCL) across datasets in four different programming languages. 

\textbf{Overall Results.} From the experimental results shown in Table~\ref{table:effectiveness}, we have two observations: \textbf{(1)~\Our~can increase the generation speed of Code LLMs significantly without compromising accuracy.} From Table~\ref{table:effectiveness}\footnote{Due to space limitations, results on overall performance under Pass@3 and Pass@5 are shown in Appendix-B.1~\cite{CodeFast}. Conclusions that hold on Pass@1 also hold for Pass@3 and Pass@5 metrics.}, we can find that our approach significantly speeds up the inference speed of all Code LLMs, with the speed-up ratio ranging from 1.34 to 4.52 times. For instance, the average generation time of \CLThirtyFour decreases from 14.51 seconds to 5.89 seconds in the MBGP dataset, which means our approach brings a 2.46 times acceleration in inference. We can also find it retains the quality of the generated code from the unchanged Pass@1 score. According to preliminary studies in \Sec~\ref{sec:preliminary}, this is because our approach can detect the excess generations and terminate the inference process early, which retains the functionality of generated code and decreases unnecessary generations. We can find that the average token length of \CLThirtyFour decrease 263.4 to 97.6 in the MBGP dataset. Besides, we evaluate the performance of \GenGuard classifiers on a manually labeled dataset\footnote{Beacause the datasets (MBPP, MBXP, etc.) only provide test cases for evaluation. We manually construct a dataset to evaluate the prediction accuracy of GenGuard.}introduced in Appendix-A.2~\cite{CodeFast} with two metrics, namely precision and recall. Experimental results show that \GenGuard obtains high precision and recall scores, both exceeding 0.95. Detailed information about the experiments is presented in Appendix-B.1~\cite{CodeFast}. \textbf{(2)~\Our 
~is effective in Code LLMs with different
parameter sizes and different programming languages.} First, from Table~\ref{table:effectiveness}, we can find that our approach works for all five Code LLMs with parameter sizes ranging from 7B to 34B. Second, our approach can be applied to the acceleration of code generation in different programming languages. We can find that in Table~\ref{table:effectiveness}, our approach is effective in code generation of four programming languages such as Python, JavaScript, Go, and C++.

\begin{figure}[t]
    \centering
    \includegraphics[ width=0.85\linewidth]{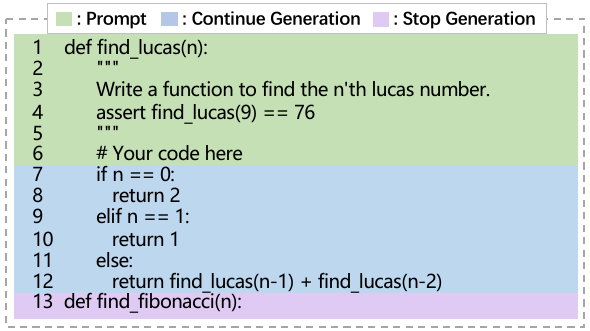}
    \figmargin
    \caption{An example of \GenGuard-enhanced Code LLM inference process.}
    \label{fig:effective_example}
    \figmargin
    \vspace{3mm}
    
\end{figure}

\textbf{Case studies.} \Fig~\ref{fig:effective_example} presents an inference example from the \GenGuard-enhanced \CLSeven. After inputting the prompt (lines 1-6), the \GenGuard-enhanced \CLSeven generate code that fulfilled requirements (lines 7-12). Following the generation of line 13, the \GenGuard module predicts that generation should stop, leading to an early termination of code generation. More cases can be found in the Appendix of replication package~\cite{CodeFast}.

\eqmargin
\begin{center}
    \begin{myboxc} \textbf{RQ1 Summary: }
    Overall, \Our~can increase the generation speed of Code LLMs significantly without compromising the quality of the generated code. Besides, our approach is effective for different Code LLMs and different programming languages.
    \end{myboxc}
\end{center}
\eqmargin

\secmargin
\subsection{RQ2: Ablation Study}
\label{sec:ablation_study_experiments}
To answer RQ2, we conduct ablation studies to investigate the impacts of the line-voting mechanism and multi-PL training data in our approach. We conduct an ablation study by removing each component at a time. Notably, when examining the impact of multi-PL training data, we train separate \GenGuard modules for Code LLMs, each using training data from a different language. For example, in our ablation study, we train a Python \GenGuard module for \CLSeven using Python training data exclusively. To assess the impact of removing the line-voting mechanism, we utilize ``token voting'', which means that we terminate the inference of Code LLMs immediately when \GenGuard predicts to stop generation. Similar to RQ1, we conduct experiments using five Code LLMs across datasets from four programming languages.

\textbf{Overall results.} From Table~\ref{tab:ablation}\footnote{Due to space limitations, results on ablation study under Pass@3, and Pass@5 are presented in Appendix-B.2~\cite{CodeFast}. Conclusions hold on Pass@1 also hold for Pass@3 and Pass@5 metrics.}, we can see that: \textbf{(1) \GenGuard modules trained with Multi-PL data have comparable performance to models trained with Mono-PL data.} 
Code LLMs enhanced with Multi-PL \GenGuard achieve similar Pass@1 scores and speed acceleration as those enhanced with Mono-PL \GenGuard module. This indicates that training a unified \GenGuard with data from multiple PLs does not compromise its predictive ability in any of these languages. 
\textbf{(2) The line-voting mechanism effectively enhances the quality of code generated by \GenGuard-enhanced Code LLMs.} Table~\ref{tab:ablation} shows that after removing the line-voting mechanism, the speed of the \GenGuard-enhanced model slightly increases, but its Pass@1 performance declines. For example, without the line-voting mechanism, the Pass@1 scores of the \GenGuard-enhanced \CLThirteen model dropped across all four datasets. This degradation can be attributed to the \GenGuard module occasionally making incorrect predictions, ending the generation process too early and producing incomplete code. In contrast, the line-voting mechanism using majority voting on the predictions of \GenGuard within a line enhances the stability of the generation results. 

\begin{figure}[t]
    \centering
    \includegraphics[ width=\linewidth]{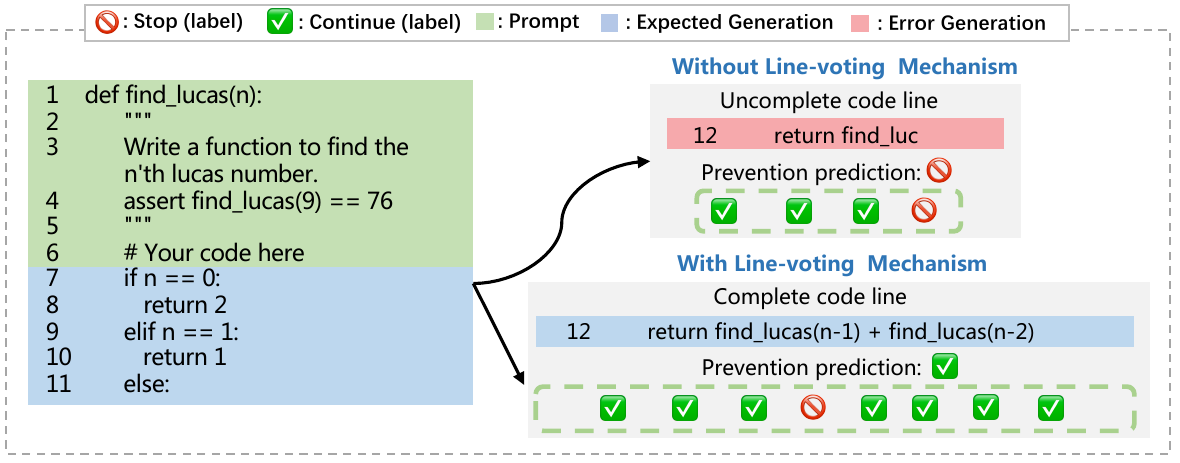}
    \figmargin
    \caption{A comparison of \GenGuard-enhanced CodeLlama-7b with \& without the line-voting mechanism.}
    \label{fig:false_example} 
\end{figure}

\textbf{Case studies.} \Fig~\ref{fig:false_example} presents an example generation with and without the line-voting mechanism. Without line-voting, the code in line 12 is incomplete. This is because the \GenGuard module makes a misjudgment when generating the token ``luc'', predicting that the current production process should stop. Consequently, the generation would be terminated at this point. However, when the line-voting mechanism is enabled, as the \GenGuard module predicts ``continue generation'' for the majority of tokens in line 12, the model then completes the generation of the entire line.

\eqmargin
\begin{center}
    \begin{myboxc} \textbf{RQ2 Summary: }
    Overall, Multi-PL \GenGuard and Mono-PL \GenGuard exhibit comparable performance. 
    And the line voting mechanism is effective and can better maintain code generation accuracy.
    \end{myboxc}
\end{center}
\eqmargin

\begin{center}
\begin{table*}[t]
\small \setlength{\tabcolsep}{1pt}
\caption{Ablation study results.}
\label{tab:ablation}
\tabmargin
\resizebox{0.95\linewidth}{!}{
\begin{tabular}{l l cc cc cc cc}\toprule
  \multirow{2}{*}{\textbf{Model}} & \multirow{2}{*}{\textbf{Type}} 
        & \multicolumn{2}{c}{\textbf{MBPP(Python)}} & \multicolumn{2}{c}{\textbf{MBJSP(Javascript)}} 
        & \multicolumn{2}{c}{\textbf{MBGP(Go)}} & \multicolumn{2}{c}{\textbf{MBCPP(C++)}} \\  
  \cmidrule(r){3-4} \cmidrule(r){5-6} \cmidrule(r){7-8} \cmidrule(r){9-10} 
 & & \textbf{Pass@1} & \textbf{Speedup} & \textbf{Pass@1} & \textbf{Speedup} & 
   \textbf{Pass@1} & \textbf{Speedup} &  \textbf{Pass@1} & \textbf{Speedup}  \\ \midrule
\multirow{3}{*}{\CLSeven} 
  & \Our        & {41.0} & $\times$4.52 & {47.2} & $\times$2.62 & {33.4} & $\times$2.07 & {43.0} & $\times$2.66 \\
  & \quad w/o MultiPL & {41.2} & $\times$4.67 & {47.2} & $\times$2.60 & {32.6} & $\times$2.16 & {43.0} & $\times$2.72  \\
  & \quad w/o LineVoting & {41.0} & $\times$5.81 & {46.2} & $\times$3.49 & {33.4} & $\times$2.29 & {42.8} & $\times$2.78 \\ \hline
\multirow{3}{*}{\CLThirteen}
  & \Our & {44.8} & $\times$3.98 & {51.6} & $\times$2.77 & {38.4} & $\times$2.73 & {51.4} & $\times$2.93 \\ 
  & \quad w/o MultiPL & {44.8} & $\times$4.03 & {51.6} & $\times$2.76 & {38.4} & $\times$2.71 & {51.4} & $\times$2.93 \\ 
  & \quad w/o LineVoting & {42.8} & $\times$5.41 & {49.4} & $\times$3.83 & {38.0} & $\times$3.09 & {49.8} & $\times$3.16 \\ \hline
\multirow{3}{*}{\CLThirtyFour} 
  & \Our & {51.6} & $\times$4.05 & {57.4} & $\times$1.50 & {42.4} & $\times$2.46 & {55.2} & $\times$2.04 \\ 
  & \quad w/o MultiPL & {51.4} & $\times$4.08 & {57.4} & $\times$1.49 & {42.4} & $\times$2.46 & {55.2} & $\times$2.02 \\ 
  & \quad w/o LineVoting & {50.8} & $\times$5.13 & {56.4} & $\times$2.04 & {42.2} & $\times$2.69 & {55.2} & $\times$2.21 \\ \hline
\multirow{3}{*}{\StarCoder} 
  & \Our & {42.6} & $\times$2.59 & {32.8} & $\times$4.11 & {31.2} & $\times$2.49 & {45.2} & $\times$2.41 \\ 
  & \quad w/o MultiPL & {42.6} & $\times$2.59 & {32.8} & $\times$4.11 & {31.2} & $\times$2.09 & {45.2} & $\times$2.41 \\ 
  & \quad w/o LineVoting & {39.2} & $\times$6.21 & {32.4} & $\times$4.89 & {30.8} & $\times$2.87 & {45.0} & $\times$2.68 \\ \hline
\multirow{3}{*}{\PhindCL} 
 & \Our & {55.8} & $\times$1.34 & {60.0} & $\times$1.44 & {42.2} & $\times$2.46 & {60.4} & $\times$1.96 \\ 
 & \quad w/o MultiPL & {55.8} & $\times$1.33 & {60.0} & $\times$1.41 & {42.2} & $\times$2.47 & {60.4} & $\times$1.94 \\ 
 & \quad w/o LineVoting & {53.4} & $\times$1.72 & {58.8} & $\times$1.92 & {41.6} & $\times$2.94 & {59.2} & $\times$2.16  \\ \bottomrule
\end{tabular}
}
\end{table*}
\end{center}

\secmargin
\vspace{-2mm}
\subsection{RQ3: Stability}
\label{sec:stability_experiments}
In RQ3, we investigate the stability of our approach under different parameter settings. We primarily explore the impact of two parameters on the performance of \Our. The first parameter is \textit{max\_new\_tokens}, as mentioned in \Sec~\ref{sec:background_inference}, which represents the maximum token sequence generation length for the Code LLM. The second parameter is the \textit{stop\_threshold} $\theta_{stop}$, indicating the probability threshold at which the \GenGuard module predicts to stop generation. Due to the space limit, we only show the experiment results of the MBPP dataset in \Fig~\ref{fig:stability}. Additional experimental results for MBGP, MBJSP, and MBCPP are provided in Appendix-B.3~\cite{CodeFast}.

Firstly, as illustrated in \Fig~\ref{fig:stability}\footnote{Due to space limitations, results on stability analysis under Pass@3 and Pass@5 are presented in Appendix-B.3~\cite{CodeFast}. Conclusions hold on Pass@1 also hold for Pass@3 and Pass@5 metrics.}, it is observed that the inference time for all \GenGuard-enhanced Code LLMs tends to be stable with the increase of \textit{max\_new\_tokens}. In contrast, the inference time for baselines significantly increases as \textit{max\_new\_tokens} rises. Notably, our approach significantly decreases inference time compared to the baseline at the same \textit{max\_new\_tokens} values. Furthermore, it is observed that our method maintains a Pass@1 rate comparable to baselines under different \textit{max\_new\_tokens} values. This indicates that our method maintains code quality consistently as \textit{max\_new\_tokens} varies. Secondly, our approach demonstrates stable performance in both inference speed and Pass@1 across various stop threshold values $\theta_{stop}$. This is because \GenGuard is highly confident in stopping prediction and can tolerate a wide range of $\theta_{stop}$ values. We analyze the stopping prediction probability distribution of \GenGuard. The results indicate that the stopping probability is distributed between 0.9 \revise{and} 1 when predicting true excess tokens and between 0  \revise{and}  0.1 when predicting true expected tokens. The experimental results are in Appendix-B.3~\cite{CodeFast}. 

\begin{figure}[t]
\centering
\begin{subfigure}{.5\linewidth}
  \centering
  \includegraphics[width=\linewidth]{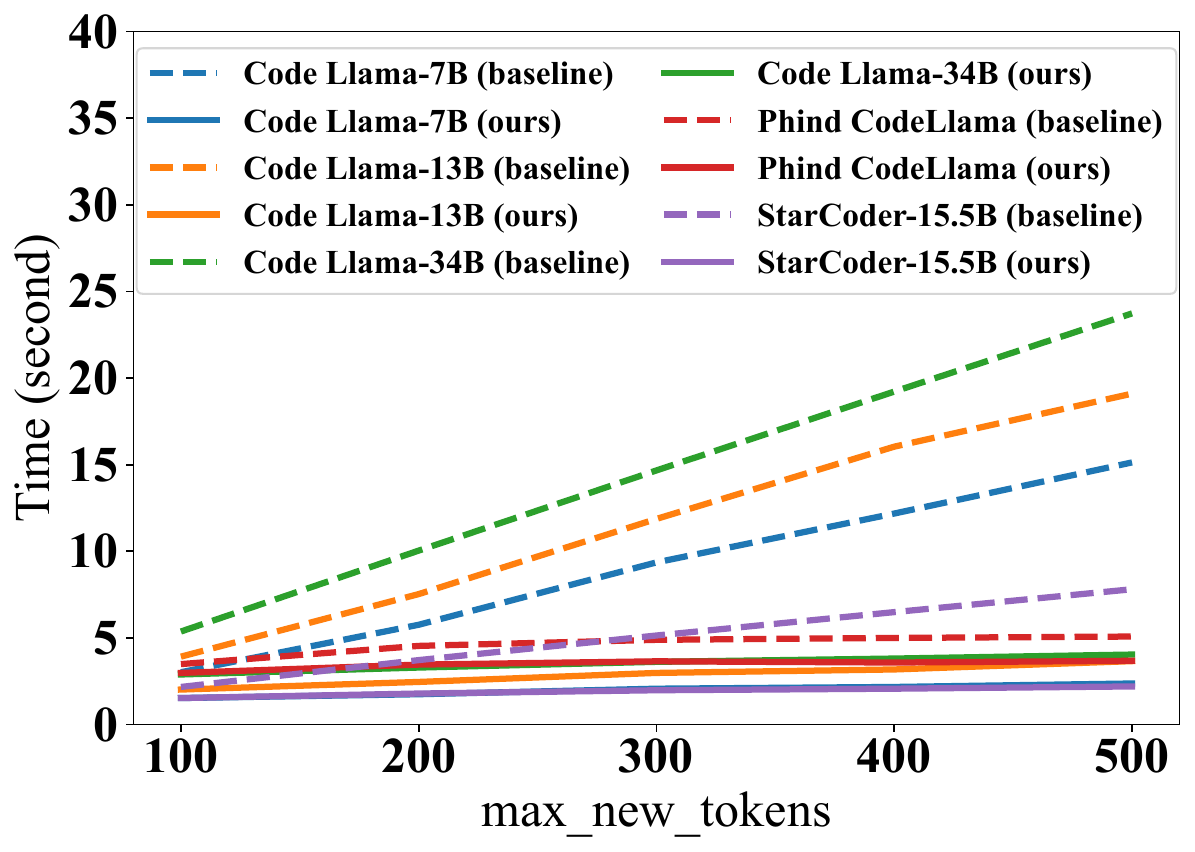}
  \tabmargin
  \label{fig:max_new_token_time}
\end{subfigure}%
\begin{subfigure}{.5\linewidth}
  \centering
  \includegraphics[width=1.0\linewidth]{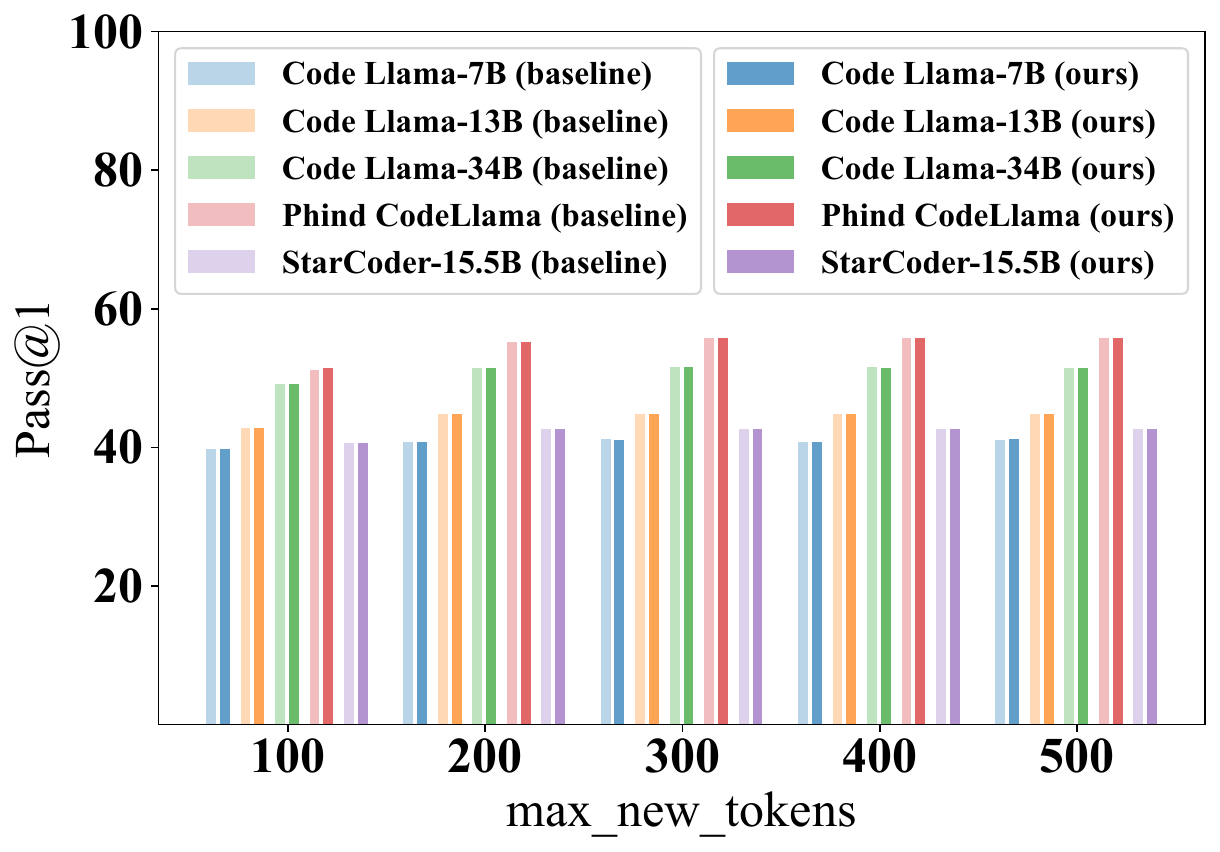}
  \tabmargin
  \label{fig:max_new_token_pass}
\end{subfigure}
\newline
\begin{subfigure}{.492\linewidth}
  \centering
  \includegraphics[width=\linewidth]{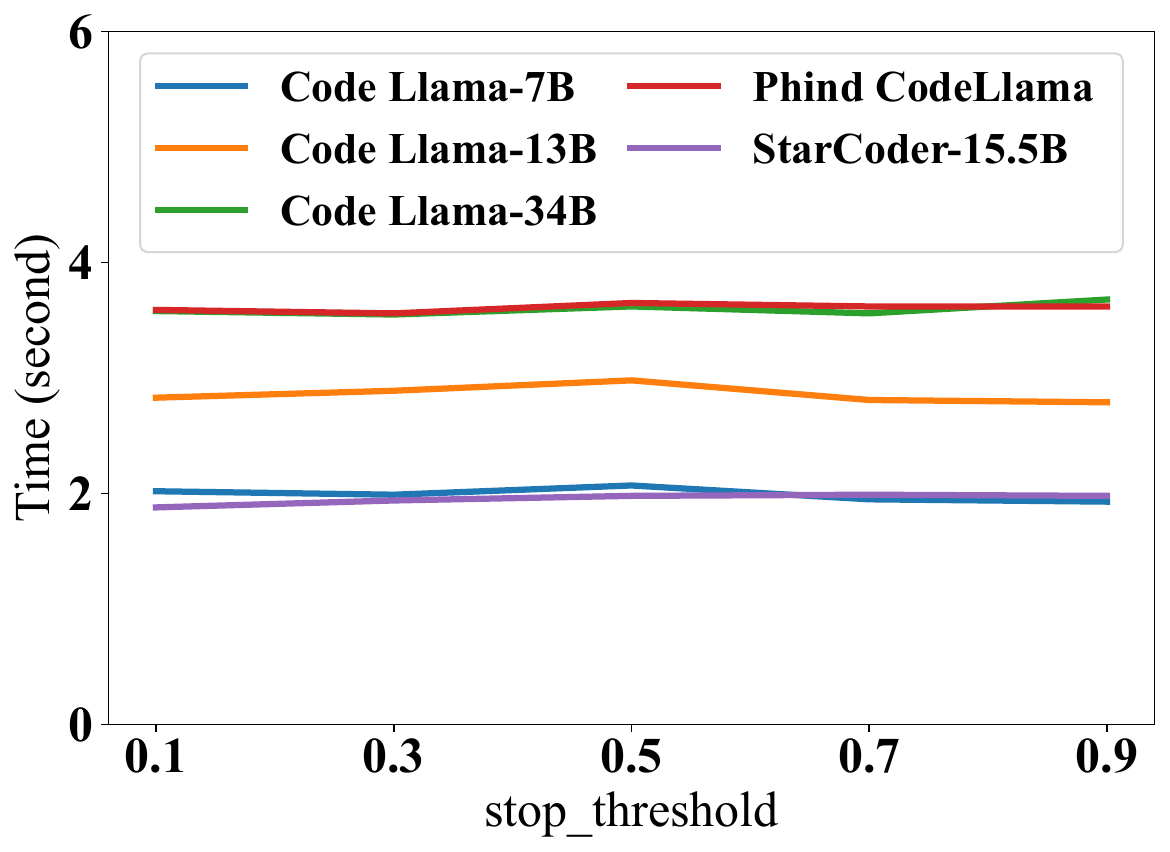}
  
  \tabmargin
  \label{fig:threshold_time}
\end{subfigure}%
\begin{subfigure}{.5\linewidth}
  \centering
  \includegraphics[width=1.0\linewidth]{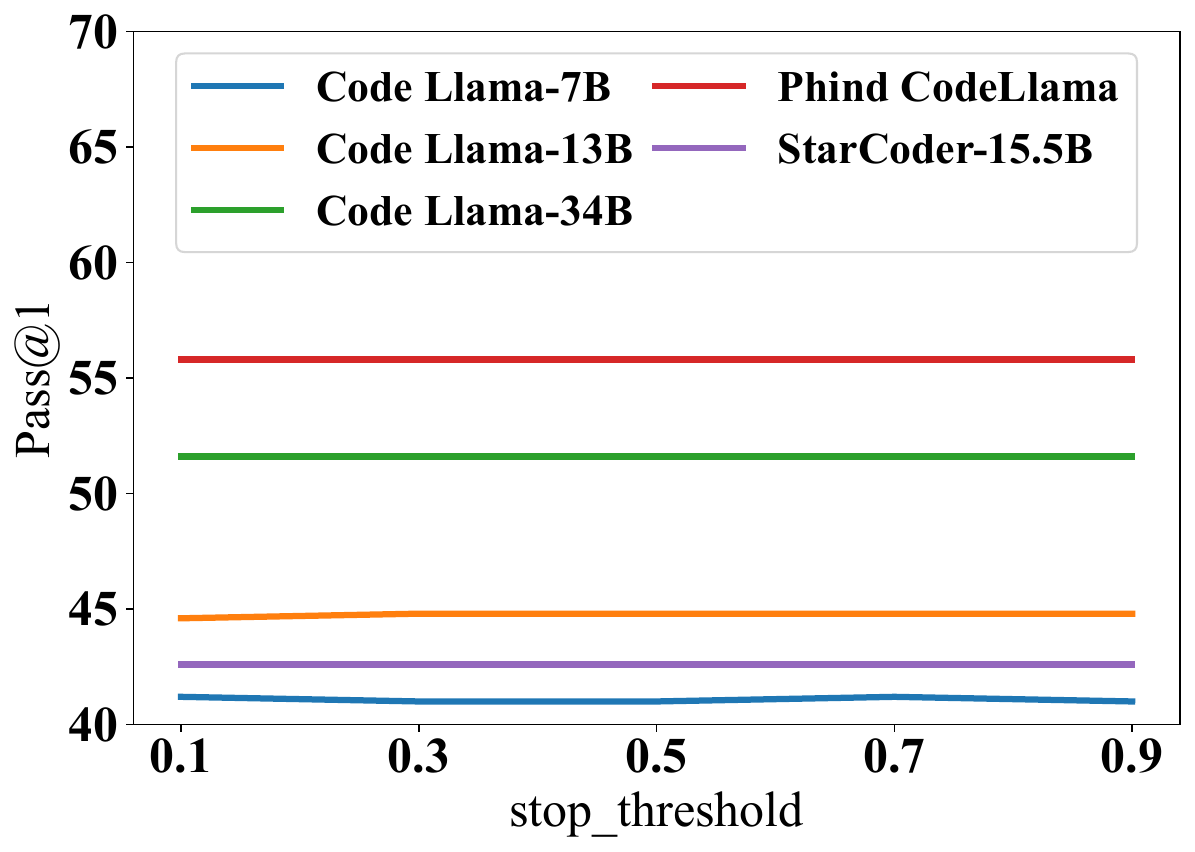}
  
  \tabmargin
  
  \label{fig:threshold_pass}
\end{subfigure}
\figmargin
\caption{Stability analysis of \Our in MBPP.}
\label{fig:stability}
\figmargin
\vspace{3mm}
\end{figure}

\eqmargin

\begin{center}
    \begin{myboxc} \textbf{RQ3 Summary: }
    Overall, \Our exhibits great stability across different parameter settings.
    \end{myboxc}
\end{center}
\eqmargin

\secmargin
\subsection{RQ4: Generalizability}
\label{sec:experiments:genralizability}
To further validate the generalizability of \Our, we evaluate the performance of five \GenGuard-enhanced models using three datasets in different PLs (Python, JavaScript, Go) from the multilingual HumanEval benchmark. 
Our models are evaluated on the test sets of these datasets without any further training. The experimental results are shown in Table~\ref{tab:general_simplified}\footnote{Due to space limitations, results on generalizability experiments under Pass@3 and Pass@5 are presented in Appendix-B.4~\cite{CodeFast}. Conclusions hold on Pass@1 also hold for Pass@3 and Pass@5 metrics.}. More detailed results with time consumption and average lengths can be found in Appendix-B.4 of replication package~\cite{CodeFast}.
We can observe that after being enhanced by the \GenGuard module, the generation speed of all five Code LLMs increases across three programming languages. Notably, this improvement is achieved while maintaining unchanged Pass@1 scores, indicating that \Our exhibits strong generalization capabilities, achieving robust performance on untrained datasets. 

\begin{center}
    \begin{myboxc}\textbf{RQ4 Summary: }
    Overall, \Our has robust performance on untrained datasets, exhibiting great generalization capabilities.
    \end{myboxc}
\end{center}

\begin{table}[t]
\centering \small
\setlength{\tabcolsep}{1pt}
\caption{Performance of \Our in untrained datasets. Python, JavaScript, and Go are short for HumanEval, HumanEval-JavaScript, and HumanEval-Go, respectively.}
\label{tab:general_simplified}
\tabmargin
\resizebox{\linewidth}{!} {
\begin{tabular}{lccccccc}\toprule
  \multirow{2}{*}{\textbf{Model}} & \multicolumn{2}{c}{\textbf{Python}} & \multicolumn{2}{c}{\textbf{JavaScript}} & \multicolumn{2}{c}{\textbf{Go}} \\ \cline{2-3} \cline{4-5} \cline{6-7} 
  & \textbf{P@1} & \textbf{Speedup} & \textbf{P@1} & \textbf{Speedup} & \textbf{P@1} & \textbf{Speedup} \\ \midrule
\CLSeven &     34.1 & $\times$1.00    & 32.9 & $\times$1.00    & 15.0 & $\times$1.00    \\  
\rowcolor{gray!15}
\textbf{\CLSevenOurs} & 34.1 & \textbf{$\times$2.71} & 32.9 & \textbf{$\times$1.99} & 15.0 & \textbf{$\times$2.11} \\ \hline
\CLThirteen & 35.4 & $\times$1.00     & 39.8 & $\times$1.00    & 20.0 & $\times$1.00    \\ 
\rowcolor{gray!15}
\textbf{\CLThirteenOurs} &35.4& \textbf{$\times$2.75} & 39.8 & \textbf{$\times$2.48} & 20.0 & \textbf{$\times$2.53} \\ \hline
\CLThirtyFour & 48.8 & $\times$1.00   & 44.1 & $\times$1.00    & 22.5 & $\times$1.00    \\ 
\rowcolor{gray!15}
\textbf{\CLThirtyFourOurs} &48.6& \textbf{$\times$2.43} &44.1& \textbf{$\times$1.38} & 22.5 & \textbf{$\times$2.17} \\ \hline
\StarCoder & 33.5 & $\times$1.00    & 15.5 & $\times$1.00    & 15.0 & $\times$1.00    \\ 
\rowcolor{gray!15}
\textbf{\StarCoderOurs} & 33.5 & \textbf{$\times$1.18} &15.5 & \textbf{$\times$5.01} & 15.0 & \textbf{$\times$2.21} \\ \hline
\PhindCL & 70.7 & $\times$1.00    & 68.9 & $\times$1.00    & 34.3 & $\times$1.00    \\ 
\rowcolor{gray!15}
\textbf{\PhindCLOurs} &70.3 & \textbf{$\times$1.48} &68.9 & \textbf{$\times$1.19} & 34.3 & \textbf{$\times$1.87} \\ \bottomrule
\end{tabular}
}
\end{table}

\vspace{-1mm}
\subsection{RQ5: Application to other scenarios}\label{sec:class}

To explore whether  \Our can be applied to other scenarios, we evaluate it on a class-level code generation dataset named ClassEval~\cite{du2023classeval}. It includes 100 class-level Python code generation tasks. Each task takes a class skeleton as input, which includes both class-level information (import statements, class name, etc.) and method-level information (method signature, functional description, etc.)~\cite{du2023classeval}. The model is asked to complete this class skeleton by infilling the implementation of unfinished methods. In our experiments, we utilize method-by-method generation strategies, which are recommended by ClassEval to evaluate models that are not GPT-4 and GPT-3.5~\cite{du2023classeval}. These strategies mean that the model generates the class method-by-method until the whole class is completed. These strategies include incremental generation strategy and compositional generation strategy~\cite{du2023classeval}. We utilize both of them for evaluation \revise{and find issues of excess generation when generating methods for the given class}. Detailed information about ClassEval and generation strategies can be found in Appendix-A.1~\cite{CodeFast}.

\textbf{Overall results.}  The experimental results with the incremental generation strategy are presented in Table~\ref{tab:classeval}. We can see that \Our effectively increases the generation speed for all five Code LLMs on ClassEval, while the
performance does not change across the three metrics (Pass@1, Pass@3 and Pass@5). 
Due to space limitations, we present experimental results with the compositional generation strategy in Appendix-B.5~\cite{CodeFast}, and results also demonstrate the efficiency of  \Our.

\begin{table}[h]
\centering \small
\setlength{\tabcolsep}{1pt}
\caption{Performance of \Our on ClassEval with incremental generation strategy. P@k represents the probability of top k-generated classes that successfully pass unit tests.}
\label{tab:classeval}
\tabmargin

\resizebox{0.9\linewidth}{!} {
\begin{tabular}{lccccc}\toprule
  \textbf{Model}
  & \textbf{P@1}  & \textbf{P@3} & \textbf{P@5} & \textbf{Time} & \textbf{Speedup} \\ \midrule
\CLSeven &     16.0   & 18.4 & 20.0  & 39.5 & $\times$1.00 \\  
\rowcolor{gray!15}
\textbf{\CLSevenOurs} & 16.0 &   18.4 & 20.0 &  18.9 &\textbf{$\times$2.09}\\ \hline
\CLThirteen & 16.0     & 21.6  & 24.0 & 50.0 & $\times$1.00  \\ 
\rowcolor{gray!15}
\textbf{\CLThirteenOurs} &16.0 & 21.6  & 24.0 & 19.1 & \textbf{$\times$2.62} \\ \hline
\CLThirtyFour & 26.0   & 28.7& 31.0 & 70.8  & $\times$1.00   \\ 
\rowcolor{gray!15}
\textbf{\CLThirtyFourOurs} &26.0& 28.7& 31.0 & 33.2 & \textbf{$\times$2.13}   \\ \hline
\StarCoder & 18.0    & 20.8 & 22.0 & 38.6 & $\times$1.00   \\ 
\rowcolor{gray!15}
\textbf{\StarCoderOurs} & 18.0 &20.8 & 22.0 & 22.2 & \textbf{$\times$1.74}  \\ \hline
\PhindCL & 18.0     & 24.9 & 26.0 & 57.1 & $\times$1.00  \\ 
\rowcolor{gray!15}
\textbf{\PhindCLOurs} &18.0  &24.9 & 26.0  & 34.6 & \textbf{$\times$1.65} \\ \bottomrule
\end{tabular}
}
\end{table}
\begin{center}
    \begin{myboxc}\textbf{RQ5 Summary: }
    Our approach can be applied to class-level code generation scenarios and effectively increases generation speed.
    
    \end{myboxc}
\end{center}

\secmargin
\section{Related Work}
\label{sec:related_work}

\vspace{1mm}
\secmargin

\subsection{LLM-based Code Generation}
Code generation has been extensively studied in recent years~\cite{le2020deep,wang2021code,hayati2018retrieval,ling2016latent} and currently the paradigm has been shifted to LLM-based code generation~\cite{zheng2024towards,wang2024RLCoder,guo2022unixcoder,luo2023wizardcoder,jain2023llm,yadav2023exploring,li2023large,liu2023refining,li2023structured,zhang2023planning}.
Some works~\cite{phind2023codellama,luo2023wizardcoder,jain2023llm,yadav2023exploring} boost the code generation ability 
through Supervised Fine-Tuning (SFT). For instance, \PhindCL~\cite{phind2023codellama} achieves superior performance over GPT-4 through SFT on high-quality code datasets based on the \CLThirtyFour model. Besides, some works~\cite{li2023large,liu2023refining,li2023structured,zhang2023planning,wei2023magicoder} propose effective prompt techniques to enhance the code generation ability of Code LLMs. For instance, Li et al.~\cite{li2023towards} propose AceCoder, which retrieves programs related to given requirements to create a prompt, enabling the model to learn from these examples and generate high-quality code. Other works~\cite{zhu2023improving,ni2023lever}  propose decoding strategies for Code LLMs to improve the performance of code generation. For example, Zhu et al.~\cite{zhu2023improving} propose AdapT sampling to improve the performance of code generation through adaptive adjusting the decoding temperature.

\secmargin
\subsection{Efficient Inference of LLMs}

Recently, many studies attempt to improve the inference efficiency of LLMs. Some works~\cite{kwon2023efficient,dao2022flashattention,dao2023flashattention2} increase the inference speed of LLMs by optimizing memory management and data access. 
Dao et al.~\cite{dao2022flashattention,dao2023flashattention2} propose FlashAttention to accelerate inference by reorganizing attention computation through a tiling approach, which significantly decreases GPU memory read/write operations. 
Other works~\cite{leviathan2023fast,zhang2023draft} accelerate the inference of LLMs by decreasing the computational time of predicting each token. 
Leviathan et al.~\cite{leviathan2023fast} propose speculative decoding, which employs a small model to predict each token and a large model to verify it. 
This approach not only enhances the generation speed but also ensures the quality of the outputs. Besides, many works attempt to increase the inference efficiency of LLMs in code intelligence tasks~\cite{sun2023don,sun2024neural,sun2024ai,gu2022accelerating}. For example,
Sun et al.~\cite{sun2023don} propose an efficient inference approach for Code LLMs. This method utilizes a Transformer-based estimator to assess the quality of prompts and prevent the completion of low-quality prompts in advance. Compared with these works, \Our improves inference efficiency for Code LLMs by preventing the generation of excess tokens. Our approach is complementary to existing efficient inference approaches, allowing for 
the synergy to further enhance inference efficiency for Code LLMs.
However, these current technologies, while effective at enhancing the inference speed of large language models, have not been tailored specifically to the unique characteristics of code generation tasks.

\secmargin
\section{Discussion}
\indent \textbf{Comparison with relevant techniques.} Recently, Sun et al.~\cite{sun2024neural} propose an effective and similar method named SEC to accelerate code completion speed by training a classifier to prevent \textit{erroneous generation}. Moreover, some bruteforce tricks specific for Python such as terminating generation upon ``\textbackslash ndef'' may easily remove excess generation. 

Compared with SEC, our approach exhibits several key differences. First,  the SEC is trained to classify \textit{erroneous generation} rather than classify \textit{excess generation}. This difference stems from the fact that the training code used by SEC does not include \textit{excess generation}, and the labeling method can not label ``stop'' at excess token, preventing SEC from learning to eliminate \textit{excess generation}. The experiment results~\footnote{\revise{We compare SEC and \Our using the MBPP dataset.}} in Appendix-D.1~\cite{CodeFast} show that our approach can increase generation speed while unchanging the model performance. 
Second, the classifier of \Our is designed and evaluated for multiple PLs, while the SEC classifier is intended for mono-PL. Third, \Our proposes an effective line-voting mechanism, which can effectively increase the accuracy of generated code as shown in experiments of RQ2. Fourth, \Our offers a more lightweight training process while SEC requires training extra intermediate LM head for each Transformer layer. 

Compared with bruteforce tricks, our approach demonstrates several distinct advantages. First, experimental results in Appendix-D.2~\cite{CodeFast} demonstrate that \Our can significantly increase the generation speed. Second, our method can be applicable to multiple programming languages, while these bruteforce tricks are for a specific programming language (Python). Third, \Our introduces an automatic training method, without the need for manually collecting specific patterns such as ``\textbackslash ndef''. Detailed discussion and experiments about relevant techniques are presented in Appendix-D~\cite{CodeFast}. 

\textbf{Application scope and limitations. } \Our aims to accelerate code generation by preventing the excess generation. 
Our study primarily focuses on function-level and class-level code generation, where we have conducted extensive experiments and demonstrated the efficiency of \Our. In the future, we will investigate the excess generation issue in more complicated code generation scenarios and evaluate \Our more comprehensively in other scenarios such as file-level and repository-level code generation.

\secmargin
\section{Conclusion}\label{sec:conclusion}
\vspace{-1mm}
In this paper, we propose a straightforward and effective approach \Our for accelerating the code generation speed of Code LLMs and show its effectiveness through extensive experiments. In our preliminary studies, we present the issue of excess generations and show that this issue is a significant factor limiting code generation speed. Then, we propose an effective inference acceleration approach \Our, which terminates the inference process early when detecting excess generations with \GenGuard module. Additionally, we propose a data construction framework to obtain training data for the \GenGuard automatically. Experimental results show that \Our effectively increases the code generation speed of Code LLMs without compromising the quality of generated code. We believe \Our can effectively improve the efficiency of Code LLMs in real-world scenarios and be utilized in IDE in the future. 
Our code and data are available at 
\textbf{\url{https://github.com/DeepSoftwareAnalytics/CodeFast}}.

\sloppy
\begin{acks}
\revise{The work is supported by CCF-Huawei Populus Grove Fund CCFHuaweiSE202301 and the National Key Research and Development Program of China (2023YFB2704801).}
\end{acks}

\normalem

\bibliographystyle{ACM-Reference-Format}
\bibliography{ref}


\end{document}